\documentclass[10pt]{iopart}

\usepackage{acronym}
\usepackage{amssymb}
\usepackage[usenames]{color}
\usepackage{hyperref}
\usepackage{graphicx}
\usepackage{textcomp}
\usepackage[sort&compress,numbers]{natbib}

\newcommand{\dd}{\mathrm{d}}
\newcommand{\sun}{\odot}

\begin{document}

\title{Turbulence in Core-Collapse Supernovae}

\author{David~Radice$^{1,2}$,
Ernazar~Abdikamalov$^3$,
Christian~D.~Ott$^4$,
Philipp~M\"osta$^{5}\footnote{NASA Einstein Fellow}$,
Sean~M.~Couch$^{6,7,8}$,
and Luke~F.~Roberts$^{6,8}$.
}
\address{$^1$ School of Natural Sciences, Institute for Advanced Study,
1 Einstein Drive, Princeton, NJ 08540, USA}
\address{$^2$ Department of Astrophysical Sciences, 4 Ivy Lane, Princeton
University, Princeton, NJ 08544, USA}
\address{$^3$ Department of Physics, School of Science and Technology,
Nazarbayev University, Astana 010000, Kazakhstan}
\address{$^4$ TAPIR, Walter Burke Institute for Theoretical Physics,
Mailcode 350-17, California Institute of Technology, Pasadena, CA 91125,
USA}
\address{$^5$ Department of Astronomy, 501 Campbell Hall \#3411,
University of California at Berkeley, Berkeley, California 94720, USA}
\address{$^6$ Department of Physics and Astronomy, Michigan State
University, East Lansing, MI 48824, USA}
\address{$^7$ Department of Computational Mathematics, Science, and
Engineering, Michigan State University, East Lansing, MI 48824, USA}
\address{$^8$ National Superconducting Cyclotron Laboratory, Michigan
State University, East Lansing, MI 48824, USA}

\begin{abstract}
Multidimensional simulations show that non-radial, turbulent, fluid
motion is a fundamental component of the core-collapse supernova (CCSN)
explosion mechanism. Neutrino-driven convection, the standing accretion
shock instability, and relic-perturbations from advanced stages of
nuclear burning can all impact the outcome of core collapse in a
qualitative and quantitative way. Here, we review the current
understanding of these phenomena and their role in the explosion of
massive stars. We also discuss the role of protoneutron star convection
and of magnetic fields in the context of the delayed neutrino mechanism.
\end{abstract}

\section{Introduction}
The formation of an iron core at the center of evolved massive stars,
i.e., with zero-age main-sequence masses of ${\sim} 10\, M_\sun$ or more,
is a robust prediction of stellar evolution theory. It is also well
established that, once it reaches a critical mass of ${\sim}1.4\,
M_\sun$, the iron core becomes dynamically unstable and succumbs to its
self-gravity. The collapse proceeds until the center reaches nuclear
density $\simeq 2.7 \times 10^{14}\ {\rm g}\ {\rm cm}^{-3}$. Then, the
\ac{EOS} stiffens because of the repulsive component of the strong
nuclear force. The inner part of the iron core, that is in sonic contact
during the collapse, halts its contraction and bounces back. A
\ac{PNS}\acused{NS} is thus formed. The outer part of the iron core, that
is infalling supersonically, rams into it producing a strong shock wave.

In the case of the most common variety of \acp{CCSN}, this shock wave is
expected to expand toward the stellar surface, depositing energy and
unbinding material, ultimately powering the explosion.  However, this
process is far from straightforward. Within ${\sim} 20\ {\rm ms}$ of its
formation, the shock wave loses about two Bethes ($1 {\rm B} \equiv
10^{51}\ {\rm erg}$) of energy due to the neutrino shock-breakout burst.
Subsequently, as the shock moves out in mass, it loses energy at the rate
of $8.8\ {\rm MeV}$ per nucleon (${\sim}1.7\ {\rm B}$ per $0.1\ M_\odot$)
as it dissociates iron-group nuclei into free nucleons. Eventually,
within a few tens of milliseconds, the shock stalls and turns into a
standing accretion shock stagnating at a radius between $100$ and $200$
kilometers. For the explosion to be successful, some physical mechanism
must revive the shock within a timescale of ${\sim}1$ second. Otherwise,
the \ac{PNS} would collapse to form a black hole.  Understanding this
mechanism is one of the most pressing problems in astrophysics.

\begin{figure}
  \begin{center}
    \includegraphics[width=0.7\textwidth]{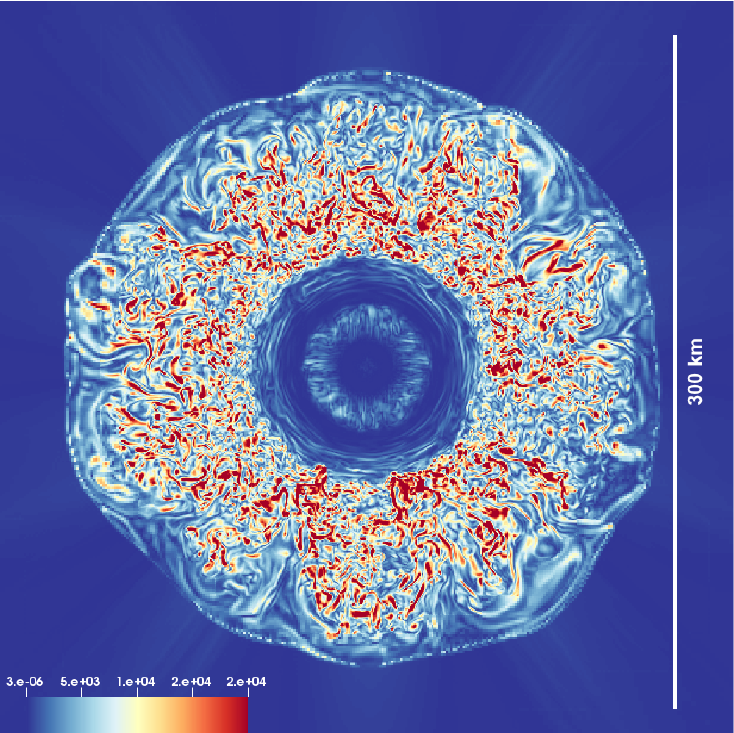}
  \end{center}
  \caption{Regions of intense vorticity during the stalled-shock phase of
  the explosion of a CCSN.  The color code shows the amplitude of the
  vorticity. The data are taken at $210$ milliseconds after bounce for a
  20-$M_\odot$ progenitor. The progenitor model was computed with MESA
  \cite{paxton:11, paxton:13}, while the simulation was performed in 3D
  using the FLASH code with M1 neutrino transport \cite{fryxell:00,
  dubey:09, oconnor:15b}. Turbulent convection immediately behind the
  shock is clearly visible and is characterized by the presence of small
  tubular structures with intense vorticity. Similar structures are known
  to characterize incompressible turbulence in periodic domains
  \cite{vincent:91}. The vortical structures near the core are due to PNS
  convection.}
  \label{fig:vort2d}
\end{figure}

The leading theoretical model for shock revival is the so-called delayed
neutrino mechanism \cite{colgate:69, bethe:85}. According to this
mechanism, it is the absorption of neutrinos radiated from the \ac{PNS}
and its cooling mantle that revives the shock. All modern simulations
show the formation of a layer behind the shock, the so-called gain
region, with net neutrino heating. Whether this is the mechanism solely
responsible for the explosion is still uncertain. For instance, we now
know that for most progenitors the delayed neutrino mechanism does not
work in spherical symmetry \cite{ramppjanka:00, liebendoerfer:01b,
thompson:03, liebendoerfer:05, sumiyoshi:05, radice:17b}. However,
successful explosions have been obtained in multiple dimensions thanks to
the presence of hydrodynamical instabilities, such as the \ac{SASI}
\cite{blondin:03, foglizzo:07} and neutrino-driven convection
\cite{herant:95, bhf:95, janka:96, foglizzo:06}. These instabilities
break the spherical symmetry of the problem and generate turbulence.
These two effects, asphericities of the flow and turbulence, are now
recognized as being crucial for successful explosions. Indeed, as shown
in Fig.~\ref{fig:vort2d}, turbulence is present over most of the region
interior to the shock.

The kinematic viscosity in the region immediately below the shock is
dominated by neutron scattering and is very small \cite{abdikamalov:15},
implying an extremely high Reynolds-number flow ($\mathrm{Re} \approx
10^{17}$; \cite{abdikamalov:15}). An important ramification for
radiation-hydrodynamics CCSN simulations is that they need to be 3D. They
also need to have sufficiently high resolution to capture the turbulent
flow according to the implicit large eddy simulation (ILES) paradigm;
whereby the bulk of the large-scale energy-containing eddies are well
resolved and numerical dissipation is used as an effective turbulent
model at small scales \cite{garnier:09, aubard:13, radice:16a}. In 1D
(spherical symmetry) there is only radial flow. In 2D (axisymmetry),
turbulence exhibits an inverse cascade that transfers kinetic energy to
large scales \cite{kraichnan:67} where it can artificially help the
explosion \cite{lentz:15, couch:15a, couch:14a}.

The purpose of this article is to review our current understanding of the
role of turbulence and other fluid instabilities in the explosion
mechanism of \acp{CCSN}. We refer to the recent reviews by
\cite{janka:12b, burrows:13a, janka:16a, mueller:16b} for more broad
overviews of \ac{CCSN} theory. The rest of this article is organized as
follows. In Section \ref{sec:conv}, we focus on the development of
neutrino-driven convection and its role in the explosion mechanism.
Section \ref{sec:sasi} is dedicated to the \ac{SASI}. In Section
\ref{sec:presn}, we consider the evolution of perturbations, originating
in convective nuclear burning shells, as they are advected and amplified
during core collapse. We discuss in more detail the role that pre-SN
perturbations can have in aiding the explosion in Section
\ref{sec:turbulence.shock}. Section \ref{sec:pns} is dedicated to the
role of \ac{PNS} convection. We review the possible role of \ac{MHD}
effects for the explosion of regular \acp{CCSN} in Section \ref{sec:mhd}.
Finally, we conclude in Section \ref{sec:conclusions} with a brief
summary and a discussion of future prospectives in the study of
turbulence in \acp{CCSN}.

\section{Neutrino-Driven Convection}
\label{sec:conv}
\subsection{The Development of Convection}
\label{sec:conv-dev}

Because of the geometric dilution of the neutrino radiation field, the net
heating rate per baryon has a maximum within the gain layer and decreases
outwards. This naturally leads to the formation of an unstable (negative)
entropy gradient. The gain region is unstable to convection according to
both the Schwarzschild \cite{chandra:39} and the Ledoux \cite{ledoux:47}
criteria.

According to the Ledoux criterion, in the case of a static background,
the growth rate of small convective eddies is given by the imaginary part
of the Brunt-V\"ais\"al\"a frequency:
\begin{equation}\label{eq:omega.bv}
  \Omega_{\rm BV}^2 =
  \frac{g}{\rho}
  \left(\frac{\partial\rho}{\partial p}\right)_{s,Y_e}
  \left[ \left(\frac{\partial p}{\partial s}\right)_{\rho, Y_e}
    \frac{\partial s}{\partial r} +
    \left(\frac{\partial p}{\partial Y_e}\right)_{\rho,s}
    \frac{\partial Y_e}{\partial r}
  \right]\,,
\end{equation}
where the electron fraction $Y_e$ is used to express the composition of
the flow, $s$ is its entropy per baryon, and $g$ is the gravitational
acceleration.

Classically, as soon as $\Omega_{\rm BV}^2 < 0$ convection is expected to
develop. However, as pointed out by Foglizzo et al.~\cite{foglizzo:06},
in the case of the gain layer in \acp{CCSN}, there is an additional
condition that needs to be satisfied for convection to develop: the
growth rate of the convective instability must be large enough to ensure
that buoyant plumes can grow sufficiently before being advected through
the gain region. This condition is expressed in terms of the
non-dimensional parameter
\begin{equation}\label{eq:foglizzo.chi}
  \chi = \int \frac{|\Omega_{\rm BV}|}{|v_r|} \dd r\,,
\end{equation}
where the integral is restricted to those regions of the gain layer where
$\Omega^2_{\rm BV}$ is negative. $\chi$ can be interpreted as the number
of e-folding times experienced by a buoyant perturbation before being
accreted into the cooling layer above the \ac{PNS}. Many numerical
simulations have shown that, when $\chi \gtrsim 3$, convection develops
\cite{foglizzo:06, scheck:08, burrows:12, hanke:12, hanke:13,
fernandez:14, couch:14a, iwakami:14, abdikamalov:15, fernandez:15a,
cardall:15}. On the other hand, the interpretation of $\chi$ in the
presence of large deviations of the background flow from spherical
symmetry is less clear. First, since Eq.~(\ref{eq:foglizzo.chi}) is
non-linear, computing $\chi$ using angle and time averaged flow variables
gives results that can be significantly different from its direct
evaluation as a volume integral \cite{fernandez:14}. This is because, in
this case, the value of $\chi$ can be dominated by points with small
radial velocities. Differences are also likely to arise depending on
whether $\chi$ is computed using the average Brunt-V\"ais\"al\"a
frequency, e.g.,~\cite{fernandez:14}, or the Brunt-V\"ais\"al\"a
frequency of the averaged flow, e.g.,~\cite{iwakami:14}. Second,
numerical simulations suggest that in the presence of sufficiently large
perturbations convection might develop also for small values of $\chi$
\cite{scheck:08, ott:13a}. Finally, once convection is fully developed,
$\chi$ often decreases to below the critical value of three
\cite{fernandez:14, couch:14a}.

\begin{figure}
  \begin{center}
    \includegraphics[width=0.49\textwidth]{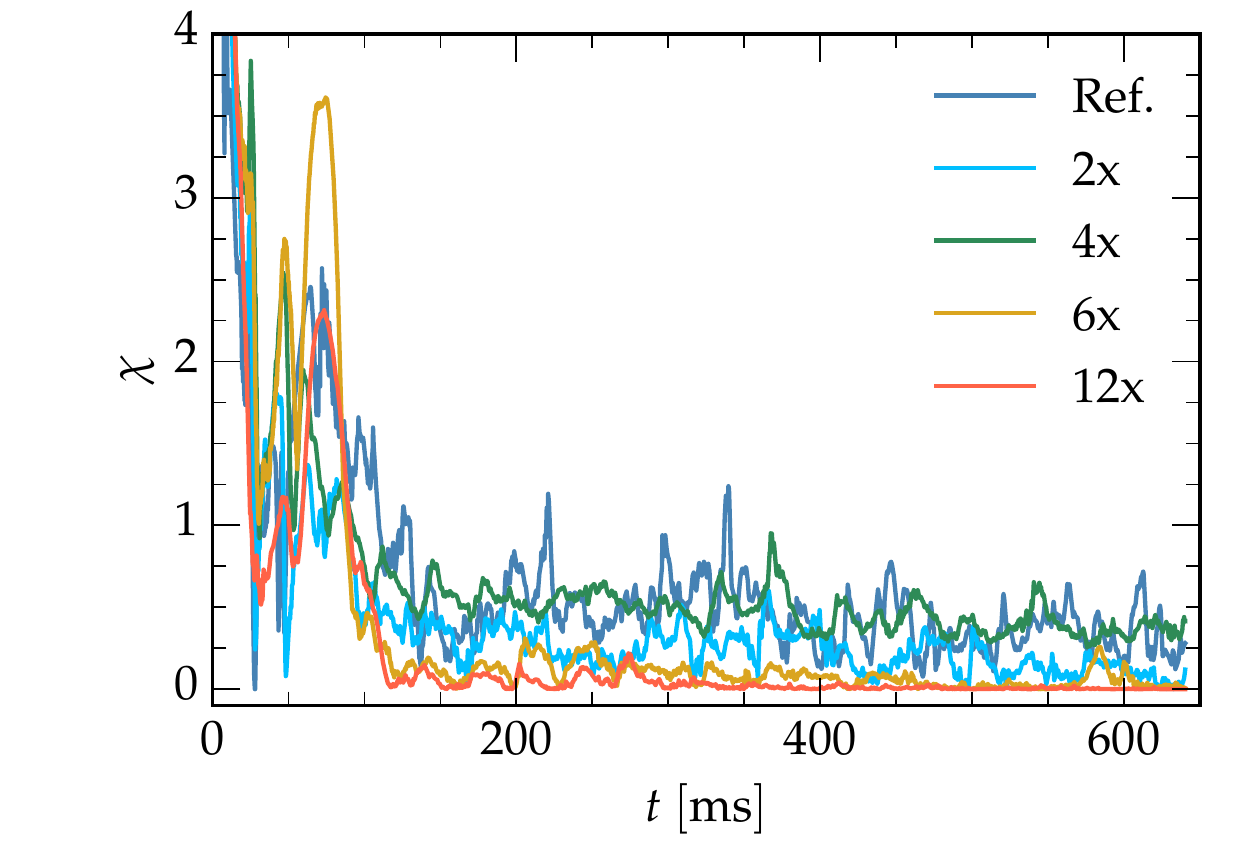}
    \hfill
    \includegraphics[width=0.49\textwidth]{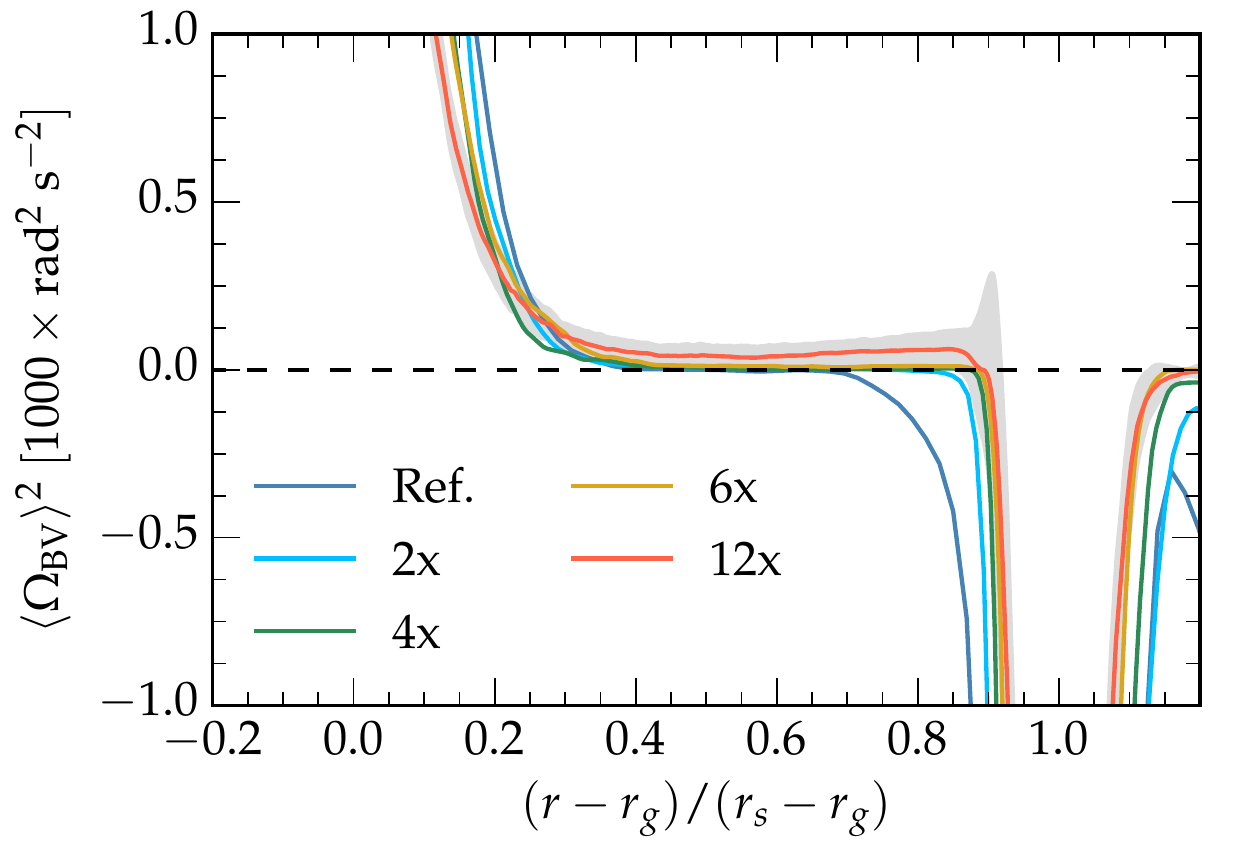}
  \end{center}
  \caption{Left panel: evolution of Foglizzo's $\chi$,
  Eq.~(\ref{eq:foglizzo.chi}). Right panel: time and angle averaged
  Brunt-V\"ais\"al\"a frequency $\Omega_{\rm BV}$ as a function of a
  normalized radial coordinate. The results are from the 3D
  high-resolution, semi-global, parameterized, neutrino-driven convection
  simulations presented in \cite{radice:16a}. Here, $\chi$ is computed
  from the angle-averaged $\Omega_{\rm BV}$ and velocity, $r_g$ is the
  radius at which the angle-averaged net neutrino heating rate first
  becomes positive, and $r_s$ is the shock radius. The initial value of
  $\chi$ is $5.33$, but it drops significantly once convection is fully
  developed and $\langle\Omega_{\rm BV}\rangle^2$ is positive over most
  of the gain region. The final value of $\chi$ also decreases with
  resolution.}
  \label{fig:foglizzo.chi}
\end{figure}

For example, Fig.~\ref{fig:foglizzo.chi} shows that if $\chi$ is computed
from the angle averaged $\Omega_{\rm BV}$ and radial velocity, then the
development of convection tends to drive $\chi$ toward zero. In
particular, the left panel Fig.~\ref{fig:foglizzo.chi} shows the
evolution of $\chi$ from a series of parametrized semi-global neutrino
driven convection simulations \cite{radice:16a}. The reference resolution
is $\Delta r \simeq 3.8\, \rm{km}$ and $\Delta \phi = \Delta \theta =
1.8^\circ$. The other simulations have their resolution increased (grid
spacing decreased) by factors $2$, $4$, $6$, and $12$. The initial value
of $\chi$ is $5.33$, but, as soon as convection is fully developed,
$\chi$ drops significantly and it reaches essentially zero in the highest
resolution runs. Despite the vanishing value of $\chi$, convection is
fully developed for the entire duration of the simulations. However, the
flow re-arranges so that $\langle \Omega_{\rm BV} \rangle^2 \gtrsim 0$
almost everywhere in the gain layer (right panel of
Fig.~\ref{fig:foglizzo.chi}). The only exceptions are the points where
the angle averaging crosses the shock surface and $\langle \Omega_{\rm
BV} \rangle$ is not particularly meaningful (these points are not
considered when computing $\chi$ from Eq.~[\ref{eq:foglizzo.chi}]).

Clearly, the interpretation of Foglizzo's $\chi$ once the flow has broken
the spherical symmetry is difficult and its value is not necessarily
indicative of whether convection or \ac{SASI} is dominating. On the
other hand, the value of $\chi$ shortly after bounce is predictive of the
onset of convection. Foglizzo's criterion also shows that convection
should become dominant for models with low accretion rates and/or high
neutrino luminosities. This has been extensively confirmed by many 3D
simulations, e.g.,~\cite{burrows:12, dolence:13, murphy:13, ott:13a,
couch:13b, couch:14a, takiwaki:14a, iwakami:14, abdikamalov:15,
melson:15a, lentz:15, melson:15b, roberts:16c}.

\subsection{The Impact of Convection}
\label{sec:conv.impact}
The importance of convection in the gain layer below the shock has been
recognized for many years, see e.g.,~\cite{bethe:90}. The results from
some of the earliest multi-dimensional (2D) simulations suggested that
convection might be strong enough to trap energy into the gain region
\cite{herant:94, herant:95}. According to these simulations, the
temperature of the gas below the shock would then continuously increase
due to neutrino absorption until a critical threshold after which the
explosion would set in. This was the so-called convective engine model.
However, the results from more advanced simulations, e.g.,~starting from
\cite{bhf:95, janka:96, mezzacappa:98}, questioned these earlier studies
and showed that the impact of convection is important, but more subtle.

Modern simulations show that once convection develops, a new equilibrium
is typically reached with no net accumulation of mass or energy in the
gain region. Instead, the explosion, if successful, is likely to develop
in a more nuanced way if the right conditions, often expressed in terms
of neutrino energy deposition and accretion rate \cite{burrows:93}, are
met. In other words, neutrino-driven convection is not able to completely
trap material in the gain region and create a run-away accumulation of
energy. However, it can aid the explosion by creating more favorable
conditions for the shock to run away, for instance, when the accretion
rate suddenly drops as compositional interfaces cross the shock.

One way convection aids the explosion is by entraining some of the gas,
which, as a consequence, has more time to absorb neutrinos \cite{bhf:95,
murphy:09, fernandez:09b, dolence:13}. Equivalently, convection
decreases the typical advection timescale
\begin{equation}\label{eq:tadv}
  t_{\rm adv} = \frac{M_{\rm gain}}{\dot{M}}\,,
\end{equation}
bringing it closer to the timescale needed to heat the material in the
gain region sufficiently to unbind it \cite{marek:09, fernandez:12,
mueller:12a}.

Another possibility is that if convection is sufficiently strong, it
could drive a redistribution of entropy leading to conditions that are
more favorable for explosion~\cite{bethe:90}. This effect was quantified
by Yamasaki \& Yamada~\cite{yamasaki:07} in terms of its reduction on the
critical neutrino luminosity \cite{burrows:93}, the neutrino luminosity
beyond which the shock is able to overcome the ram pressure of the
accreting material. They showed that if convection is able to eliminate
the radial gradient of the angle-averaged specific entropy, then the
critical luminosity is effectively decreases by a factor of two. On the
other hand, simulations have shown that, in conditions typical of
\acp{CCSN}, neutrino-driven convection is not sufficiently vigorous to
realize the scenario hypothesized by Yamasaki \& Yamada~\cite{murphy:08,
murphy:11, hanke:12, dolence:13, ott:13a, radice:16a} and the reduction
in the critical neutrino luminosity is smaller than what they predicted.

Neutrino-driven convection behind the shock is turbulent and the chaotic
motion of the fluid can result in an effective additional pressure
support behind the stalled shock. This was first pointed out by Murphy et
al.~\cite{murphy:13} and later confirmed and extended by Couch \& Ott
\cite{couch:15a} and by Radice et al.~\cite{radice:16a}. Following the
original argument by Murphy et al.~\cite{murphy:13}, we consider the
Rankine-Hugoniot momentum condition for a standing accretion shock,
\begin{equation}\label{eq:rankine.hugoniot}
  \rho_d v_d^2 + p_d = \rho_u v_u^2 + p_u\,,
\end{equation}
where $\rho$ is the density, $v$ the radial velocity, and $p$ the
pressure. We have also denoted with $\cdot_d$ and $\cdot_u$ the
downstream and upstream quantities respectively.

\begin{figure}
  \begin{center}
    \includegraphics[width=0.49\textwidth]{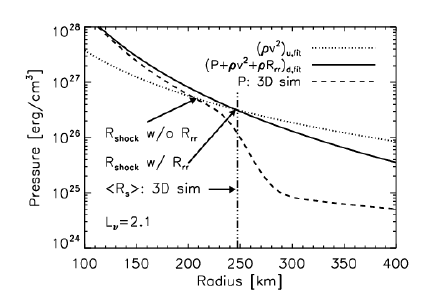}
    \hfill
    \includegraphics[width=0.49\textwidth]{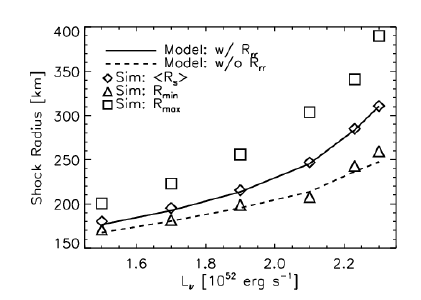}
  \end{center}
  \caption{Left panel: power-law fits of the upstream and downstream
  components of Eq.~(\ref{eq:rankine.hugoniot.turb}). The location of the
  average shock radius $\langle R_s \rangle$ and of the predictions from
  Eq.~(\ref{eq:rankine.hugoniot}), which does not account for the
  turbulent pressure, and from Eq.~(\ref{eq:rankine.hugoniot.turb}),
  which includes the effect of turbulent pressure, are also indicated.
  Right panel: minimum, maximum, and average shock radii in parametrized
  3D simulations of neutrino-driven convection with different neutrino
  luminosities. Also shown are the predictions from
  Eqs.~(\ref{eq:rankine.hugoniot}) and (\ref{eq:rankine.hugoniot.turb}).
  Reprinted from J.~Murphy, J.~Dolence, and A.~Burrows, \textit{The
  Dominance of Neutrino-Driven Convection in Core-Collapse Supernovae},
  The Astrophysical Journal \textbf{771}, 52 (2013) \cite{murphy:13},
  \textcopyright~AAS. Reproduced with permission.}
  \label{fig:rankine.hugoniot.turb}
\end{figure}

In the presence of turbulence, \cite{murphy:13} suggested to modify
equation (\ref{eq:rankine.hugoniot}) in a way akin to a Reynolds
decomposition and write it as
\begin{equation}\label{eq:rankine.hugoniot.turb}
  \rho_d v_d^2 + \rho_d R_{rr} + p_d = \rho_u v_u^2 + p_u\,,
\end{equation}
where $R_{rr}$ is the radial component of the Reynolds stress tensor,
\begin{equation}\label{eq:reynolds.tensor}
  R_{ij} = \langle \delta v_i\, \delta v_j \rangle\,,
\end{equation}
and $\delta v_i$ is the difference between the fluid velocity and the
angle-averaged radial velocity of the flow. Although not entirely
rigorous, Eq.~(\ref{eq:rankine.hugoniot.turb}) has been shown to be well
verified in numerical simulations if angular averages are used to compute
the respective quantities \cite{murphy:13, couch:15a, mueller:15,
radice:16a}; see Fig.~\ref{fig:rankine.hugoniot.turb}. In that Figure,
reproduced from \cite{murphy:13}, the shock location in 3D parametrized
neutrino-driven convection simulations is compared with the prediction
from Eq.~(\ref{eq:rankine.hugoniot.turb}). It is found that the effective
pressure support generated by the turbulence accounts for an increase in
the shock radius of ${\sim} 25\%$ \cite{murphy:13}. Furthermore, Couch \&
Ott~\cite{couch:15a} showed that the effective pressure support due to
turbulence can be as large as ${\sim} 50\%$ of the thermal pressure.

Couch \& Ott~\cite{couch:15a} also showed that the transition to
explosion in \acp{CCSN} is associated with an increase in the total
turbulent energy in the gain region, defined as
\begin{equation}
  E_{\rm turb} = \frac{1}{2} \rho |\delta \mathbf{v}|^2\,.
\end{equation}
The importance of turbulent energy can be better understood by recasting
Eq.~(\ref{eq:rankine.hugoniot.turb}), following \cite{radice:15a}
\begin{equation}\label{eq:rankine.hugoniot.turb.energy}
  \rho_d v_d^2 + (\gamma_{\rm th} - 1) \rho_d \epsilon_{\rm th} +
  (\gamma_{\rm turb} - 1) \rho_d \epsilon_{\rm turb} =
    \rho_u v_u^2 + p_u\,,
\end{equation}
where $\epsilon_{\rm th}$ is the specific internal energy downstream of
the shock, $\gamma_{\rm th} \simeq 4/3$ is the adiabatic index for a
radiation pressure dominated gas, and $\epsilon_{\rm turb} = E_{\rm
turb}/\rho_d$ is the specific turbulent energy density. $\gamma_{\rm
turb}$ is an effective adiabatic index which is equal to $5/3$ for
isotropic turbulence and to $2$ for anisotropic turbulence with $R_{rr}
\simeq R_{\theta\theta} + R_{\phi\phi}$. The latter case is commonly
realized in convection simulations, e.g.,~\cite{arnett:09, murphy:13,
couch:15a, radice:16a}, and should be the relevant case for \acp{CCSN}.
Equation (\ref{eq:rankine.hugoniot.turb.energy}) illustrates that
turbulent kinetic energy contributes to the overall pressure support
behind the shock. More importantly, since $\gamma_{\rm turb} >
\gamma_{\rm th}$, Eq.~(\ref{eq:rankine.hugoniot.turb.energy}) shows that
turbulent kinetic energy is a more ``valuable'' form of energy than
thermal energy, since it can more effectively overcome the ram pressure
ahead of the shock. The ratio $\epsilon_{\rm turb}/\epsilon_{\rm th}$ is
then a crucial quantity for the revival of the shock.

Another formulation of Eq.~(\ref{eq:rankine.hugoniot.turb.energy}) was
proposed by M\"uller \& Janka \cite{mueller:15}. Instead of the
turbulent energy, they computed the turbulent pressure from the rms Mach
number of the turbulence
\begin{equation}\label{eq:rms.mach}
  \langle \mathcal{M}^2 \rangle = \frac{2 E_{\rm kin, \theta}}{M_{\rm
  gain} c_{s,d}^2} \simeq \frac{\epsilon_{\rm turb}}{c_{s,d}^2}\,,
\end{equation}
$c_{s,d}$ being the sound speed downstream of the shock. With this
definition the total pressure support behind the shock (turbulent plus
thermal) is given by
\begin{equation}\label{eq:total.pressure}
  p = p_{\rm th} ( 1 + \gamma_{\rm th} \langle \mathcal{M}^2 \rangle )
  \simeq p_{\rm th} \left( 1 + \frac{4}{3}\langle \mathcal{M}^2 \rangle
  \right)\,.
\end{equation}

Radice et al.~\cite{radice:16a} presented a more formal derivation of
Eq.~(\ref{eq:rankine.hugoniot.turb}), which they reformulated as a global
integral condition implicitly determining the location of the shock.
Their analysis confirmed the previous interpretation of the importance of
the effective turbulent pressure. Furthermore, they found that the
non-radial fluid motion due to turbulence also contributes significantly
to the effective pressure support of the flow. This is because, even
though the angular momentum of the flow is conserved, the fluid still
experiences a net centrifugal force.

More recently, Mabanta \& Murphy~\cite{mabanta:17} considered the impact
of turbulence on the explosion conditions using a new semi-analytic
model. They extended the work of Yamasaki \& Yamada~\cite{yamasaki:07} by
combining the neutrino-driven convection model developed by Murphy \&
Meakin~\cite{murphy:11} with the steady-state model of Murphy \&
Dolence~\cite{murphy:15}. They pointed out that neutrino-driven
convection is able to tap into the gravitational potential energy
associated with the unstable stratification of the material behind the
shock. According to their model, buoyancy would transform this potential
energy into kinetic energy, which would then be transformed into thermal
energy by turbulent dissipation. Mabanta \& Murphy~\cite{mabanta:17}
argued that the subsequent enhanced thermalization of the flow would be
the most important consequence of turbulence for the explosion mechanism
of \acp{CCSN}. However, this effect has not yet been investigated using
multi-dimensional simulations, so its quantitative impact remains
unclear.

Clearly, all of the effects of neutrino-driven convection just discussed
are deeply connected. The increased residence time of fluid elements in
the gain region, the convective luminosity, the turbulent dissipation,
and the effective turbulent pressure support, can all be described using
the Reynolds decomposition, e.g.,~\cite{murphy:11, radice:16a}, of the
mass, energy, and momentum conservation equations respectively.

\subsection{Turbulent Energy Cascade}
On the basis of the previous discussion, we can conclude that the ratio
between turbulent and thermal pressure is an important parameter
affecting the dynamics of the shock revival in neutrino-driven
\acp{CCSN}. This ratio in turn is set by the efficiency with which
kinetic energy is transferred to small scales by the turbulence cascade,
where it is dissipated into heat.  Understanding how the turbulent
cascade operates in neutrino-driven convection and how well it is
captured in finite-resolution simulations is important to be able to
quantify the role of turbulence in the explosion and the fidelity of the
numerical models.

The turbulent energy spectrum is the main diagnostic used to study the
turbulent cascade \cite{endeve:12, hanke:12, couch:13b, dolence:13,
handy:14, couch:14a, couch:15a, abdikamalov:15}. Most investigators
have focused on the transverse turbulent velocity spectrum
\begin{equation}\label{eq:transverse.spectrum}
  E(\ell) = \sum_{m=-\ell}^\ell \left| \int \sqrt{\rho} v_t Y_\ell^m \dd
  \Omega \right|^2\,,
\end{equation}
where $Y_\ell^m$ are the spherical harmonics, $v_t$ is the non-radial
part of the velocity, and the integral is evaluated on a shell at a fixed
radius. Another quantity of interest is the specific turbulent energy
spectrum
\begin{equation}\label{eq:vel.spec}
  E(k) = \frac{1}{2} \sum_{i=1}^3 \int_{\mathbb{R}^3} \delta(|\mathbf{k}|
  - k) \widehat{\delta v_i^\ast}(\mathbf{k}) \widehat{\delta
  v_i}(\mathbf{k}) \dd \mathbf{k}\,,
\end{equation}
where $\widehat{\cdot}$ is the Fourier transform, defined as
\begin{equation}
  \widehat{\delta v_i}(\mathbf{k}) =
    \int_{\mathbb{R}^3} \delta v_i(\mathbf{x})
    e^{2\pi\mathrm{i}\mathbf{k}\cdot\mathbf{x}} \dd \mathbf{x}
 \end{equation}
and $\cdot^\ast$ denotes the complex conjugation. Finally, $\delta v_i$
is defined as in Eq. (\ref{eq:reynolds.tensor}). Apart from the
normalization factor, this quantity coincides with the velocity
power-spectrum, which is also the Fourier transform of the two-point
correlation function of the velocity \cite{frisch:96}.

At scales that are well separated from those at which turbulence is
driven by neutrinos and dissipated by microphysical processes, the
statistical properties of turbulence in steady state, including the
spectra, are expected to be universal~\cite{frisch:96}. At these
intermediate scales, in the so-called inertial range, the spectra $E(\ell)$
and $E(k)$ should become self-similar power laws~\cite{frisch:96}. For
example, in Kolmogorov's classical theory of turbulence $E(k) \propto
k^{\alpha}$ with index $\alpha = -5/3$. Note that, in the limit of large
$\ell$, and since $\rho \simeq {\rm const}$ at a fixed radii, $E(k)\propto
k^\alpha$ is equivalent to $E(\ell) \propto \ell^{\alpha}$, e.g.,
\cite{peebles:93}, chapter 21.

In the phenomenology associated with Kolmogorov's theory the energy flux
across scales is constant in the inertial range. Consequently, the energy
dissipation rate in well developed turbulence only depends on the large
scale dynamics, which provides the ``inflow'' boundary condition for the
energy transport to small scale. Consequently, the presence of a well
resolved inertial range is regarded as a sufficient condition to ensure
that a numerical simulation is capturing the turbulent dissipation of
energy from the largest eddies.

One possible issue is that Kolmogorov's theory of turbulence may not
necessarily apply to neutrino-driven convection. Indeed, the flow in
\acp{CCSN} is affected by nuclear reactions and neutrino irradiation,
effects that are not considered in the classical theory. Moreover,
turbulence in the postshock region of \acp{CCSN} is anisotropic, as
gravity breaks the local rotational symmetry of the fluid equations
\cite{murphy:11, murphy:13, handy:14, couch:15a, radice:16a}, mildly
compressible (reaching pre-explosion Mach numbers of $\simeq 0.3 - 0.5$
\cite{couch:13d, mueller:14}), and driven at multiple scales by buoyancy.

Nevertheless, the plasma flow in the gain region is well described using
classical hydrodynamics. This is because the plasma constituents have
characteristic cross-sections of ${\sim} 10^{-24}\, {\rm cm}^2$. Given
the typical temperatures ${\sim}1\, {\rm MeV}$ and densities ${\sim}
10^8\, {\rm g}\, {\rm cm}^{-3}$ in the gain region, this implies mean
free paths around $10^{-7}\, {\rm cm}$ and thermodynamical equilibration
timescales of few femtoseconds. These scales are many orders of magnitude
smaller than those of interest for the modeling of the explosion, so the
conditions for the validity of the hydrodynamical description are fully
realized. Moreover, momentum exchange due to neutrinos in the gain region
is unlikely to damp turbulence due to the long neutrino mean free paths
compared to the typical size of eddies in the inertial range
\cite{abdikamalov:15}.

Furthermore, anisotropic and mildly compressible flows have been shown to
follow Kolmogorov's theory in other contexts, e.g.~\cite{porter:98,
benzi:08, radice:15a}. The effect of buoyancy has been considered by
Goldreich \& Keeley~\cite{goldreich:77} who argued that Kolmogorov's
phenomenology should still apply, because turbulent stresses become
dominant over buoyancy forces at sufficiently small scales. Here, we
present a revised and generalized version of their argument.

Consider eddies with typical size $l$ (not to be confused with the
spherical harmonic index $\ell$), s.t.~$\eta \ll l \ll L$, where $L$ is
the energy injection scale, i.e., the size of the gain region, and $\eta$
is the dissipation scale. The buoyancy forces acting on these eddies
scale as the entropy fluctuations at the scale of the eddies:
\begin{equation}\label{eq:buoyancy.force}
  F^B_l \sim \delta s_l\,.
\end{equation}
Inertial forces scale as\footnote{Note that these are a manifestation of
the $(\mathbf{v}\cdot\nabla)\mathbf{v}$ term in the Euler's equations.}
\begin{equation}\label{eq:inertial.force}
  F^I_l \sim \frac{(\delta v_l)^2}{l}\,.
\end{equation}
Let us assume that velocity and entropy fluctuations scale as power laws
of $l$\footnote{Equivalently, we assume the existence of an inertial
range.}
\begin{equation}
  (\delta v_l)^2 \sim l^{2\beta}\,, \qquad
  \delta s_l \sim l^\gamma\,.
\end{equation}
Note that from Eq.~(\ref{eq:vel.spec}) and the properties of the Fourier
transform, it follows that the power-spectrum of the velocity is also a
power law with index $\alpha = -2\beta-1$, e.g., \cite{frisch:96}.  Under
these assumptions, the ratio between inertial and buoyant forces acting
on eddies with typical scale $l$ is
\begin{equation}\label{eq:inertia.vs.buoyancy}
  \frac{F^I_l}{F^B_l} \sim l^{2\beta-\gamma-1}\,,
\end{equation}
which shows that inertial forces dominate on sufficiently small
scales as long as
\begin{equation}\label{eq:inertial.dominance}
2\beta < \gamma + 1\,.
\end{equation}
This condition is verified for turbulence with Kolmogorov's spectrum,
where $\beta = \gamma = 1/3$ suggesting that a flow satisfying
Eq.~(\ref{eq:inertial.dominance}) will develop a Kolmogorov spectrum at
scales $\eta \ll l \ll l^\star$, $l^\star$ being the scale at
which $F^I_{l^\star} \sim F^B_{l^\star}$ over one eddy turnover
timescale $\tau_l = l/\delta v_l$.

\begin{figure}
  \begin{center}
    \includegraphics[width=0.5\textwidth]{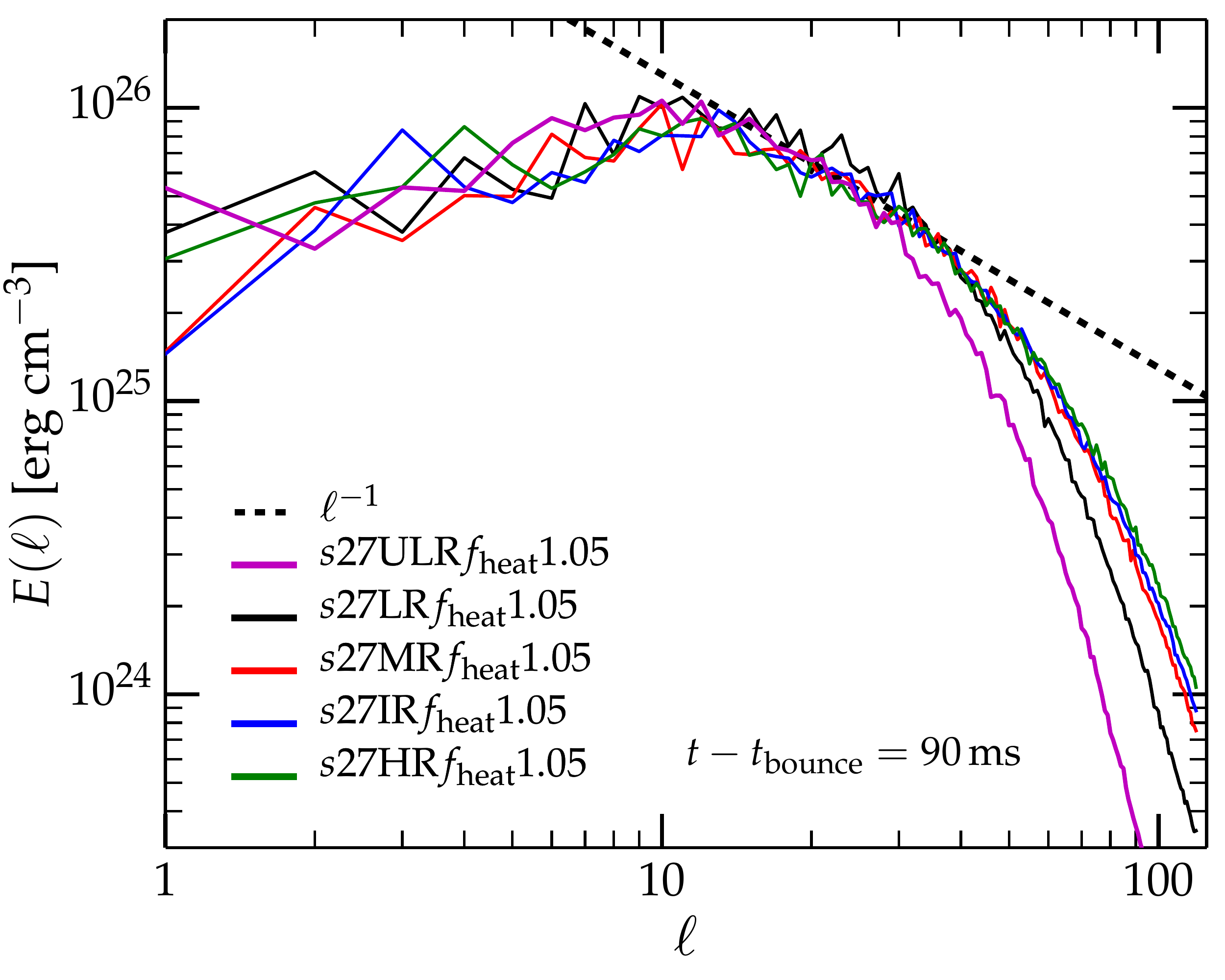}
  \end{center}
  \caption{Transverse energy spectra at different resolutions for one of
  the models from Abdikamalov et al.~\cite{abdikamalov:15}. Note that the
  spectra have a rather shallow $\ell^{-1}$ dependency at intermediate
  scales of $10 \lesssim \ell \lesssim 40$. Reprinted from
  E.~Abdikamalov, C.~D.~Ott, D.~Radice, L.~F.~Roberts, R.~Haas,
  C.~Reisswig, P.~M\"osta, H.~Klion, and E.~Schnetter,
  \textit{Neutrino-Driven Turbulent Convection and Standing Accretion
  Shock Instability in Three-Dimensional Core-Collapse Supernovae}, The
  Astrophysical Journal \textbf{808}, 70 (2015) \cite{abdikamalov:15},
  \textcopyright~AAS. Reproduced with permission.}
  \label{fig:abdikamalov.spec}
\end{figure}

We stress that Eq.~(\ref{eq:inertial.dominance}) is satisfied for most
values of $\beta$ and $\gamma$ in the physically relevant range $0 <
\beta,\gamma < 1$ (note that $\beta \geq 1$ or $\gamma \geq 1$ would
imply smooth, laminar velocity or entropy fields, respectively, e.g.,
\cite{frisch:96}). On the basis of this argument, we can conclude that
buoyancy should become negligible at small scales $l \ll l^\star$ and
deviations from Kolmogorov scaling should not be expected, unless
compressibility or other non-ideal effects become important, or
Eq.~\ref{eq:inertial.dominance} is violated.

Hanke et al.~\cite{hanke:12} claimed to observe a turbulent energy
spectrum with $\alpha \simeq -5/3$. Similar results where later reported
by Handy et al.~\cite{handy:14} and Summa et al.~\cite{summa:17}.
However, later simulations employing Cartesian grids and higher
resolutions found shallower power-spectra with $\alpha \simeq -1$
\cite{dolence:13, couch:14a, couch:15a, abdikamalov:15}. See
Fig.~\ref{fig:abdikamalov.spec} for an example from Abdikamalov et
al.~\cite{abdikamalov:15}. Several authors (e.g., \cite{dolence:13,
couch:14a}) have argued that the $\alpha = -1$ scaling observed in 3D
simulations could be due to the physical nature of the postshock
turbulent flow that deviates significantly from the assumptions of
Kolmogorov turbulence. However, this is unlikely on the basis of the
argument we presented above. Furthermore, a scaling with $\alpha = -1$ is
unphysical because it would predict infinite turbulent kinetic energy in
the limit of infinite resolution and zero viscosity.

Abdikamalov et al.~\cite{abdikamalov:15} argued that a common issue in
these studies is that the resolution requirements for the recovery of an
inertial range had been severely underestimated. Their interpretation is
corroborated by the results of local simulations in periodic cubic
domains, which showed that resolutions between $512^3$ and $1024^3$ are
needed to recover an inertial range \cite{porter:98, sytine:00,
radice:15a}. In contrast, typical global 3D \ac{CCSN} simulations have
resolutions roughly equivalent to those of a $64^3$ local simulation
\cite{abdikamalov:15}.

\begin{figure}
  \begin{center}
    \includegraphics[width=0.5\textwidth]{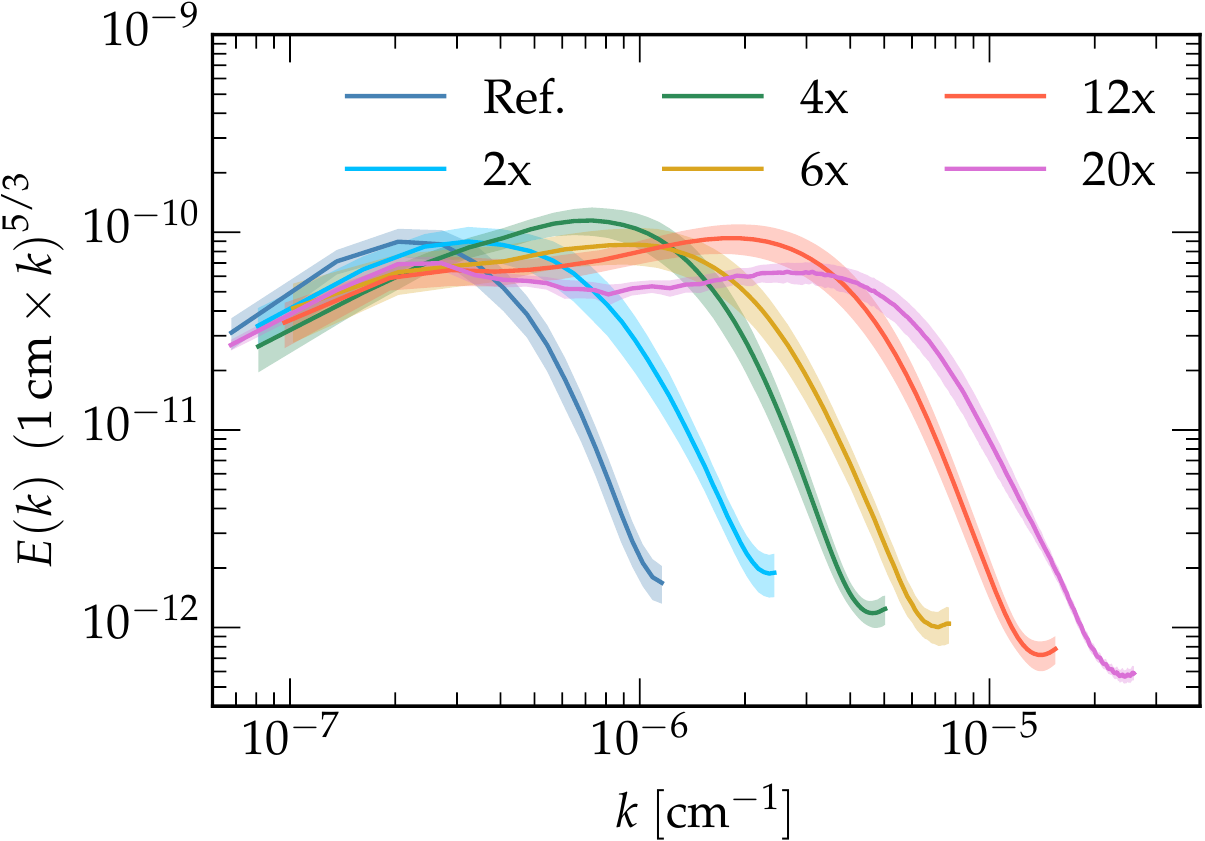}
  \end{center}
  \caption{Compensated velocity power-spectra at increasing resolutions
  in 3D, semi-global, parameterized, neutrino-driven convection
  simulations. As the resolution is increased, the turbulent specific
  kinetic energy spectrum approaches the $k^{-5/3}$ slope predicted by
  Kolmogorov's theory. Reprinted from D.~Radice, C.~D.~Ott,
  E.~Abdikamalov, S.~M.~Couch, R.~Haas, and E.~Schnetter,
  \textit{Neutrino-Driven Convection in Core-Collapse Supernovae:
  High-Resolution Simulations}, The Astrophysical Journal \textbf{820},
  76 (2016) \cite{radice:16a}. \textcopyright~AAS. Reproduced with
  permission.}
  \label{fig:radice.spec}
\end{figure}

In the light of local simulation results, Abdikamalov et
al.~\cite{abdikamalov:15} suggested that the $\ell^{-1}$ spectra found by
\cite{dolence:13, couch:14a, couch:15a, abdikamalov:15} are the result of
the so-called bottleneck effect \cite{yakhot:93, she:93, falkovich:94,
verma:07, frisch:08}. This is a phenomenon arising because of the
indirect influence of the (numerical) viscosity on the dynamics at large
scale, which results in a build up of energy at wavenumbers close to the
dissipation scale. The presence of a numerical bottleneck effect is also
likely to result in an artificially large ratio $\epsilon_{\rm
turb}/\epsilon_{\rm th}$, which would create more favorable conditions
for the explosion, as tentatively confirmed by Abdikamalov et
al.~\cite{abdikamalov:15}.

In this context, the $\ell^{-5/3}$ spectra from Hanke et
al.~\cite{hanke:12} and Handy et al~\cite{handy:14} might be interpreted
as arising from a misidentification of the dissipation range and inertial
ranges. Indeed, the spectra from Hanke et al.~\cite{hanke:12} appear to
become shallower with increasing resolution. This would be expected if
the putative inertial range with logarithmic slope $-5/3$ was actually a
part of the exponentially decaying dissipation range.

More recently, Radice et al.~\cite{radice:16a} performed a very extensive
resolution study of a parametrized neutrino-driven convection model.
Their results are reproduced in Fig.~\ref{fig:radice.spec}. They found
shallow velocity power spectra at their coarsest resolution, which was
intermediate between the medium and the high resolution of Hanke et
al.~\cite{hanke:12}. As they increased the resolution they eventually
recovered Kolmogorov scaling, but only at a resolution $\simeq 16$ times
higher than the highest resolution of Hanke et al.~\cite{hanke:12}.
These results provide additional support to the interpretation by
Abdikamalov et al.~\cite{abdikamalov:15} of the $k^{-1}$ spectra seen in
many simulations as being due to the insufficient resolution, although a
definite confirmation would require the development of techniques for the
measurement of the kinetic energy flux accross scales. See, e.g.,
\cite{radice:15a} for a study of this kind in the context of local
simulations. Radice et al.~\cite{radice:16a} also provided solid
numerical evidence for Kolmogorov's phenomenology in neutrino-driven
convection. However, the idealized simulations of Radice et
al.~\cite{radice:16a} did not include the feedback of convection on the
neutrino luminosity, which might give raise to a new phenomenology. This
limitation needs to be addressed by future studies.

\section{SASI-Generated Turbulence}
\label{sec:sasi}

\noindent The standing accretion shock instability (\ac{SASI}) was
discovered by Blondin~et al.~\cite{blondin:03}, who performed simplified
2D hydrodynamic simulations of stalled (``standing'') accretion shocks.
They found that the stalled shock develops an oscillatory instability
dominated by an (in terms of spherical harmonics) $\ell = 1$ ``sloshing''
mode. Later studies in 3D showed the prevalence of an azimuthal $m = 1$
spiral mode (e.g., \cite{blondin:07, iwakami:08, fernandez:10,
fernandez:15a, hanke:13, couch:14a, abdikamalov:15}), which is enhanced
by rotation (e.g., \cite{iwakami:09,nakamura:14,blondin:17}).

\begin{figure}
  \centering
  \includegraphics[width=0.6\textwidth]{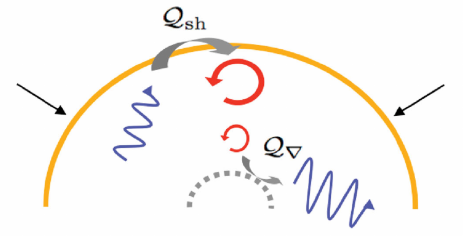}
  \label{fig:sasi}
  \caption{Schematic represenation of the advective-acoustic
    instability underlying \ac{SASI}. An entropy/vorticity
    perturbation is created at the shock front (orange half-circle) by
    the accretion flow. This perturbation is advected purely radially
    (red) to the edge of the protoneutron star (indicated by a dashed
    gray half-circle) . There, it is strongly decelerated and
    converted into an upward pressure perturbation (i.e., acoustic
    wave) that has radial and transverse components (blue). Reprinted
    from J.~Guilet \& T.~Foglizzo, \textit{On the linear growth mechanism
    driving the standing accretion shock instability}, Monthly Notices of
    the Royal Astronomical Society \textbf{421}, 546 (2012)
    \cite{guilet:12}, \textcopyright~J.~Guilet \& T.~Foglizzo. Reproduced
    with permission.}
\end{figure}

In a series of papers \cite{foglizzo:06, foglizzo:07, scheck:08,
foglizzo:09, sato:09, guilet:12, yamasaki:07}, Foglizzo and colleagues
identified SASI as an advective-acoustic instability (though see
\cite{blondin:06,laming:07} for a purely acoustic interpretation). In the
advective-acoustic scenario depicted in Fig.~\ref{fig:sasi}, a small
entropy or vorticity perturbation enters through the shock and is
advected with the subsonic accretion flow toward the \ac{PNS}.  At the
steep density gradient marking the \ac{PNS} edge, it is decelerated,
causing an outward propagating pressure (i.e., acoustic) wave with radial
and transverse components.

In the non-linear regime of \ac{SASI}, its oscillations reach large
amplitudes and cause large spatially and temporally varying expansions
and recessions of the shock front. The postshock entropy depends on the
preshock velocity in the frame of the shock. The smaller the shock
radius, the higher the preshock velocity and the higher the postshock
entropy. Hence, \ac{SASI}-driven variations in the shock radius lead to
radial entropy gradients from which turbulent convection grows
\cite{scheck:08,burrows:06,guilet:10,endeve:12}. Because of the lateral
and azimuthal \ac{SASI} shock motions, the radial preshock accretion flow
impacts the shock at an oblique angle, leading to deflection and large
non-radial velocity components. In the highly non-linear regime,
non-radial streams from different phases of a \ac{SASI} cycle collide,
forming shear-regions, secondary shocks, and supersonic accretion funnels
\cite{scheck:08, burrows:06, burrows:07a}. All these non-linear \ac{SASI}
features create conditions susceptible to turbulence and simulations show
that fully turbulent (i.e., as turbulent as permitted by the employed
resolution) postshock flow can develop from \ac{SASI} even in \ac{CCSN}
cores that are initially stable to neutrino-driven convection (e.g.,
\cite{scheck:08, endeve:12, mueller:12b, fernandez:15a}).

In 2D/3D hydrodynamic simulations, \ac{SASI} is observed to saturate in
strength/amplitude when the aforementioned non-linear flow features
appear. Guilet~et al.~\cite{guilet:10} analyzed the saturation behavior
of \ac{SASI} and proposed that saturation sets in when a parasitic
instability is able to grow fast enough to compete with \ac{SASI} by
impeding the advective-acoustic cycle. They found that both the
Rayleigh-Taylor instability (growing on entropy gradients) and the
Kelvin-Helmholtz instability (growing in shear flows) can act as such a
parasitic instability, but that the Rayleigh-Taylor instability may be
the dominant saturation agent. In any case, both parasitic instabilities
are well known to result in turbulent flow. This suggests that turbulence
is a natural consequence of \ac{SASI}, which is confirmed by simulations
(e.g., \cite{endeve:12,scheck:08,couch:14a,abdikamalov:15}).

Extensive research went into understanding the interplay of \ac{SASI} and
neutrino-driven convection, often with a focus on the question of which
of the two instabilities is dominant in \ac{CCSN}e. Because of the
advective nature of the postshock \ac{CCSN} flow, neutrino-driven
convection can grow only if seed perturbations are large enough, the
entropy gradient is steep enough, and/or the advection timescale is long
enough (cf.~Section~\ref{sec:conv-dev}, Eq.~\ref{eq:foglizzo.chi}, and
\cite{foglizzo:06,scheck:08}).  \ac{SASI} growth, on the other hand, is
fastest if the advection timescale is short \cite{foglizzo:06,scheck:08}.
\ac{CCSN} simulations with conditions in which neutrino-driven convection
grows fast and is strong and sustained do not appear to develop
large-amplitude \ac{SASI} (e.g., \cite{ott:13a, abdikamalov:15,
couch:14a, murphy:13}). In other simulations, e.g., those in which
neutrino heating is weak or that use a progenitor with a particularly
high accretion rate (e.g., \cite{hanke:13, couch:14a, abdikamalov:15}),
neutrino-driven convection may fail to develop or may be present only
temporarily. In these scenarios, \ac{SASI} eventually dominates.

While turbulence appears to be a natural consequence of both
instabilities, it remains to be clarified if SASI-induced turbulence
is different in character from neutrino-driven turbulent
convection. If it is related predominantly to entropy gradients and
the Rayleigh-Taylor instability \cite{guilet:10,scheck:08}, then its
properties may be expected to be similar to turbulence resulting from
neutrino-driven convection. This is supported by the turbulence
analyses of recent detailed 3D \ac{CCSN} simulations with \ac{SASI}
(e.g., \cite{couch:14a,abdikamalov:15}). On the other hand,
Endeve~et al.~\cite{endeve:12} found in their 3D simulations
that SASI-induced turbulence is driven predominantly by shear flows
and is partially supersonic. However, Endeve~et al.'s focus was on
turbulent magnetic field amplification and they employed an idealized
setup that left out much of the essential \ac{CCSN} microphysics. More
work in 3D with high-resolution full-physics simulations will be
needed to fully characterize SASI-induced turbulence.

\section{Pre-Supernova Turbulence}
\label{sec:presn}

\subsection{Turbulent Convection in Advanced Burning Stages}
Most \ac{CCSN} simulations start at the onset of the collapse of
perfectly spherically-symmetric cores, computed using 1D stellar
evolution codes.  These employ effective treatments of convective mixing
and overshooting, e.g., \cite{cox:68, paxton:11, paxton:13}, that can
account for some of the effects of convection, especially over secular
timescales. However there are still important uncertainties associated
with the use of these prescriptions, partly because turbulent convection
in stellar interiors is still not completely understood, see e.g.,
\cite{meakin:07b, arnett:09, viallet:13, herwig:14, woodward:15,
couch:15b, arnett:16, chatzopoulos:16, mueller:16, cristini:16,
collins:2017a}, for some recent works.

A more immediate concern for \ac{CCSN} modelers is the fact that
one-dimensional stellar evolution codes cannot predict the properties of
non-spherical perturbations of the progenitor present at the onset of
collapse. Convective velocities during advanced burning stages
immediately prior to collapse could reach hundreds of
kilometers per second \cite{couch:15b, mueller:16, collins:2017a}.
Convection, as well as gravity waves excited in the iron core during
silicon shell burning \cite{goldreich:90}, will be frozen in the
collapsing core after the onset of collapse. They will be amplified as
they cross the shock (Sec.~\ref{sec:perturbations.growth}), and they may
play an important role in triggering explosions
(Sec.~\ref{sec:turbulence.shock}).

\subsection{The Growth of Seed Perturbations in the Accretion Flow}
\label{sec:perturbations.growth}

In the process of stellar core-collapse, the convective flow in nuclear
burning shells undergoes profound evolution during its journey towards
the shock. In the absence of nuclear burning and dissipative processes,
the specific entropy of the flow is conserved, while the convective
velocities increase due to conservation of angular momentum. If present,
nuclear burning should in principle amplify convection, although the
degree of amplification is limited if the collapse timescale is shorter
than the convective turnover timescale. The dissipative processes should
do the opposite. The vorticity waves, which have solenoidal velocity
field, perturb the isodensity contours of the collapsing star, leading to
the emission of powerful pressure waves \cite{mueller:15}. Entropy
fluctuations also generate pressure waves as they accrete in a
converging flow (e.g, \cite{kovalenko:98,foglizzo:00,foglizzo:01}).

The evolution of hydrodynamic perturbations in accretion flows has
been studied in a number of works (e.g.,
\cite{kovalenko:98, lai:00, foglizzo:01, cao:10, takahashi:14, mueller:17}). A
general convective flow can be decomposed into the three physical
modes: acoustic, vorticity, and entropy waves (e.g.,
\cite{kovasznay:53}). Kovalenko \& Eremin \cite{kovalenko:98} studied
the evolution of all these three modes for Bondi accretion. Using
linearized hydrodynamics equations, they obtained
simple analytical relations in the limit $r \rightarrow 0$ for the
scaling of perturbation amplitudes with radius. In particular, for
$\ell>0$ acoustic and vorticity modes in the regime of supersonic mean
flow, they find
\begin{eqnarray}
  \frac{\delta \rho}{\rho} &\propto& 2 \ell(\ell+1) r^{-1/2}, \\
  \frac{\delta \upsilon_r}{\upsilon_r} &\propto& \ell (\ell+1) r^{(4-3\gamma)/4}, \\
  \frac{\delta \upsilon_\perp}{\upsilon_r} &\propto& r^{-1/2}, \\
  \frac{\delta \upsilon_\mathrm{rot}}{\upsilon_r} &\propto& r^{-1/2},
\end{eqnarray}
where $\gamma$ is the adiabatic index of the gas, $\rho$ and $\upsilon_r$
are unperturbed values of density and radial velocity, while $\delta
\rho$, $\upsilon_\perp$, and $\upsilon_\mathrm{rot}$ are the perturbations
of density and angular components of velocity, respectively. Lai \&
Goldreich \cite{lai:00} found similar relations for perturbations in
collapsing clouds in the same $r\rightarrow0$ limit.

Takahashi \& Yamada \cite{takahashi:14} performed an analysis of the
evolution of infalling perturbations in spherical supersonic Bondi
accretion without imposing the $r \rightarrow 0$ limit and obtained the
following scaling relations
\begin{eqnarray}
  \frac{\delta \rho}{\rho} &\propto& \ell r^{-(5-3\gamma)/4}, \\
  \frac{\delta \upsilon_r}{\upsilon_r} &\propto& \ell r^{(5-3\gamma)/4}, \\
  \frac{\delta \upsilon_\perp}{\upsilon_r} &\propto& \mathrm{const}, \\
  \frac{\delta \upsilon_\mathrm{rot}}{\upsilon_r} &\propto& \ell r^{-1/2},
\end{eqnarray}
These are somewhat different from corresponding scaling relations in
\cite{kovalenko:98, lai:00} and parts of the differences stem from the $r
\rightarrow 0$ limit imposed in these works. A common limitation shared
by these works is that they either consider only the supersonic region of
accretion \cite{lai:00,takahashi:14} or only either supersonic or
subsonic solutions of the Bondi accretion \cite{kovalenko:98}. In CCSNe,
the convective Si/O burning shells initially infall subsonically, but
then gradually accelerate and become supersonic. A detailed analysis of
the evolution of convective perturbations from the subsonic to the
supersonic regime, including the effects of possible nuclear burning and
turbulent dissipation, is currently lacking in the literature.

M\"uller et al.~\cite{mueller:17} performed supernova simulations of a 3D
$18M_\odot$ progenitor model. They observed that, prior to collapse, the
convective motion in the O shell primarily of solenoidal velocity
field, while the density perturbations remain small. However, by the
time the O shell reaches the shock, they observed strong density
perturbations with amplitude $\delta \rho/\rho \simeq 0.1$ and peak
angular wavenumber of $\ell \simeq 2$. They found that the spatial
spectrum of the density perturbations prior to shock-crossing is very
similar to that of the turbulent Mach number in the O shell prior to
collapse. More specifically, for the normalized time-averaged spectrum of
density perturbation, they found
\begin{equation}
  \frac{\hat \rho_\ell}{\hat{\rho_0}} \simeq 0.5 \hat{\cal{M}}_{r,\ell},
\end{equation}
where $\hat{\cal{M}}_{r,\ell}$ is the spectrum of the Mach number of
radial velocity in the middle of the O shell (see their equation 6). Such
as simple relation between the spectrum of the velocity perturbation in
the oxygen shell and that of density perturbations just ahead of the
shock has not been established by the aforementioned studies of the
evolution of perturbations in Bondi accretion. This discrepancy may stem
from the fact that the perturbative studies consider only the supersonic
regime of collapse and use stationary Bondi flow, which neglects the
details of progenitor density structures (see \cite{mueller:17} for
in-depth discussion).

\section{Turbulence Shock Interaction}
\label{sec:turbulence.shock}

\subsection{Numerical Results}
\label{sec:turbulence.shock.numerical}

Couch \& Ott \cite{couch:15a} were the first to show the impact of the
accretion of perturbations in simulations. They considered the
15-$M_\odot$ progenitor from Woosley \& Heger \cite{woosley:07}, which
they evolved in 3D. They performed simulations either with the original,
unperturbed, progenitor structure, or with velocity perturbations. They
found that the accretion of perturbations enhances the turbulent kinetic
energy in the gain region, especially at large scales. This has a number
of beneficial effects for the explosion, as discussed in
Sec.~\ref{sec:conv.impact}, and, in the case of the simulations of
\cite{couch:15a}, it was sufficient to trigger an explosion in an
otherwise failing model.

M\"uller \& Janka \cite{mueller:15} further explored the impact of
perturbations in the accretion flow in 2D axisymmetric simulations. They
systematically studied the effects of the character and angular scale of
the perturbations. They found that in the case of velocity
perturbations, low-order modes ($\ell = 1$ or $\ell = 2$) are the most
effective in triggering explosions. M\"uller \& Janka \cite{mueller:15}
quantified the impact of perturbations in terms of the critical neutrino
luminosity and found a decrease of up to $10\%$ for large-scale velocity
perturbations. Density perturbations were also found to be very
effective. However, a significant impact was found only for large
perturbations, corresponding to pre-collapse turbulent Mach numbers of
$\gtrsim 0.3$, which might be present only in a small fraction of
progenitors \cite{collins:2017a}.

In their analysis of the effects of perturbations on the post-shock
turbulence, M\"uller \& Janka \cite{mueller:15} emphasized the role of
the directionally-dependent modulation of the accretion rate ahead of the
shock. However, analytic studies (\cite{abdikamalov:16};
Sec.~\ref{sec:shock.turbulence.analytic}) have shown that the effects of
shock-turbulence interaction in \acp{CCSN} have a rather complex
phenomenology and that their effectiveness depends on the relative phase
of entropy and vorticity perturbations accreted through the shock.  In
any case, both simulations and analytic theory suggest that the effect of
perturbations depends on their detailed nature. This motivates the
development of well motivated perturbed progenitor models.

Chatzopoulos et al.~\cite{chatzopoulos:14} developed a systematic method
to characterize pre-collapse asphericities in \ac{CCSN} progenitors. They
proposed a method to extract characteristic properties of convection from
multi-dimensional simulations of advanced burning stages in a way that
allows them to be injected in otherwise spherical progenitor models at
the verge of core collapse. More recently, Couch et al.~\cite{couch:15b}
and M\"uller et al.~\cite{mueller:16, mueller:17} simulated, for the
first time, \ac{CCSN} progenitors that were self-consistently evolved
in multi-D in the last minutes prior to collapse.

Couch et al.~\cite{couch:15b} studied a 15-$M_\odot$ progenitor. They
self-consistently simulated the last few minutes of Si-burning, the
collapse of the iron core, and the subsequent core bounce and explosion.
They used a reduced 21-isotope nuclear network and a neutrino leakage
scheme with parametrized heating. As a baseline for comparison, they
re-simulated the collapse, bounce, and subsequent evolution of the same
iron-core after angle averaging. They found that (non-spherical)
turbulent fluctuations produced during Si burning resulted in an earlier
and more energetic explosion. A possible caveat of this first study,
however, is that Si burning was simulated only in an octant (assuming
reflection symmetry across the coordinate planes). This prevented the
appearance of convective cells with large angular scale, which seem to be
the most effective for triggering explosions \cite{mueller:15}.
Consequently, the effects of perturbations might be even larger than
found by Couch et al.~\cite{couch:15b}.

M\"uller et al.~\cite{mueller:16, mueller:17} considered instead an
18-$M_\odot$ progenitor. They simulated the last ${\sim}5$ minutes of
evolution of the oxygen-burning shell in $4\pi$ (but excised most of the
Fe/Si core of the star) \cite{mueller:16}. The resulting progenitor was
subsequently evolved in 3D with their fast multi-group transport method
\cite{mueller:15}. They found that perturbations determined the outcome
of core collapse: while the progenitor with perturbations exploded
${\sim}0.3$ seconds after bounce, the spherically-averaged progenitor did
not explode within the end of their simulation (0.625 seconds after
bounce). A third progenitor, obtained with an artificially reduced
burning rate resulting in smaller convective velocities prior to
collapse, exploded at ${\sim}0.5$ seconds after bounce.

\subsection{Analytical Results}
\label{sec:shock.turbulence.analytic}

\subsubsection{Linear Analysis of Shock-Turbulence Interaction.}
Abdikamalov et al.~\cite{abdikamalov:16} studied the impact of pre-shock
perturbations on the evolution of the shock and post-shock flow using a
linear perturbation theory known as the linear interaction analysis
(LIA) (e.g., \cite{sagaut:08}). In the LIA, the flow is decomposed into
mean and fluctuating parts. The unperturbed shock is assumed to be
planar and the mean flow is assumed to be uniform in space. While
these are approximations, the strength of the LIA lies it its
simplicity, which allows one to obtain deep insight into the physics
of shock-turbulence interaction (e.g., \cite{sagaut:08}). Furthermore, in
its regime of applicability, when the turbulent Mach number of the
perturbations ahead of the shock is small, the LIA shows excellent
agreement with high-resolution numerical simulations \cite{ryu:14}.

In the LIA, the mean flow obeys the usual Rankine-Hugoniot conditions,
while the fluctuations obey the linearized version of the jump
conditions. The upstream flow completely determines the downstream flow
via these conditions (e.g.,~\cite{mahesh:96, wouchuk:09, huete:12,
huete:13, huete:17}). The perturbations can be decomposed into vorticity,
entropy, and acoustic modes, each of which evolve independently at the
linear order \cite{kovasznay:53}. With this decomposition, the LIA
formalism can be applied to each mode individually. The results can then
be integrated to get the full depiction of shock-turbulence interaction
to linear order.

\begin{figure}
  \begin{center}
    \includegraphics[width=0.5\textwidth]{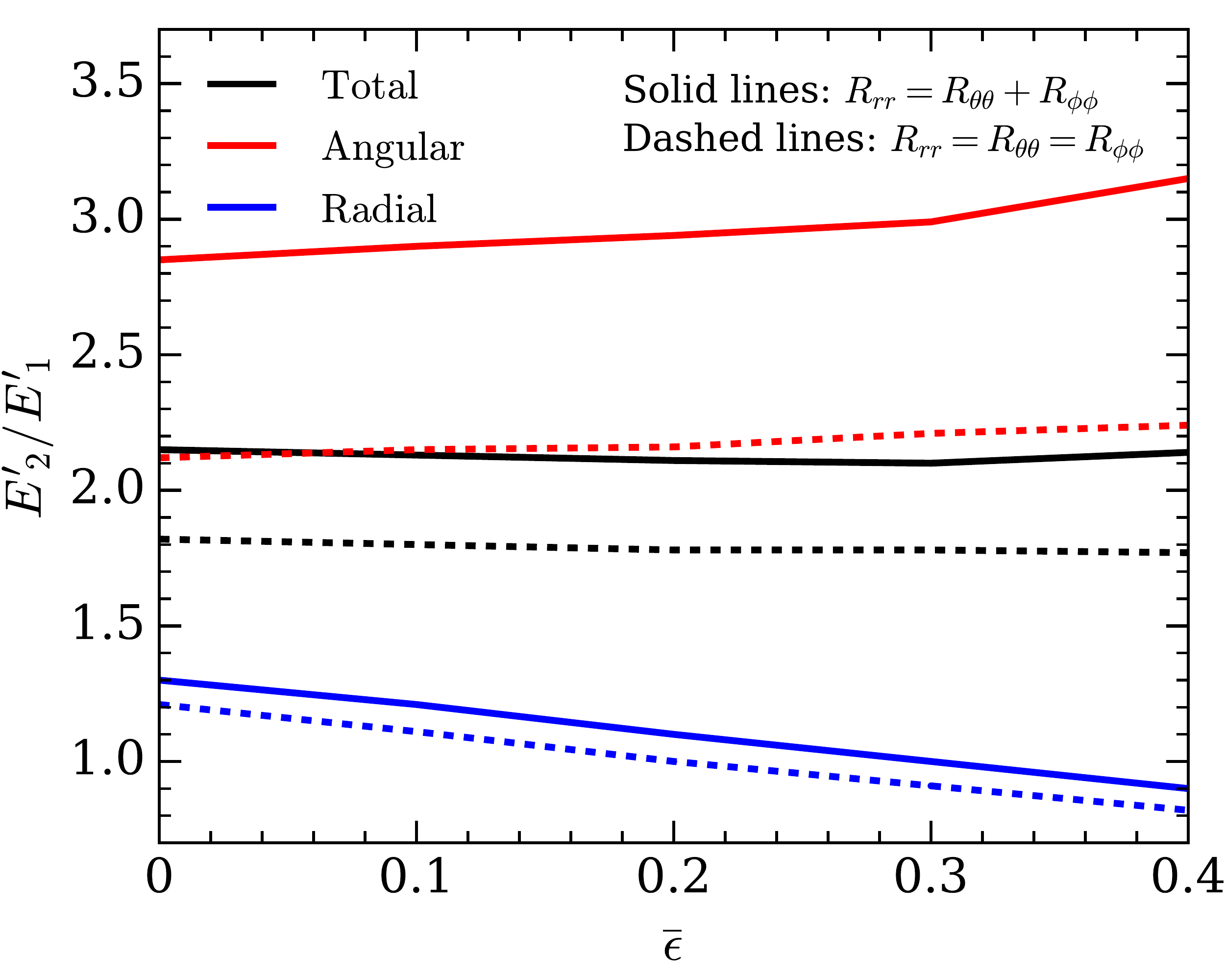}
  \end{center}
  \caption{The amplification of turbulent kinetic energy at shock
  crossing for incoming incident waves as a function of dimensionless
  nuclear dissociation parameter. The latter characterizes the strength
  of nuclear dissociation at the shock, larger values of $\bar\epsilon$
  corresponding to stronger dissociation~\cite{fernandez:09a,
  fernandez:09b}. The black lines show the amplification of the total
  kinetic energy, while blue and red lines show the amplifications of
  radial and angular components of kinetic energy. The solid lines
  represent anisotropic turbulence characterized by the relation
  $R_{rr}=R_{\theta\theta}+R_{\phi\phi}$, while the dashed lines
  represent fully isotropic turbulence. The Mach number of the pre-shock
  mean flow is assumed to be $5$. Reprinted from E.~Abdikamalov,
  A.~Zhaksylykov, D.~Radice, and S.~Berdibek, \textit{Shock-turbulence
  interaction in core-collapse supernovae}, Monthly Notices of the Royal
  Astronomical Society \textbf{461}, 3864 (2016) \cite{abdikamalov:16},
  \textcopyright~E.~Abdikamalov, A.~Zhaksylykov, D.~Radice, and
  S.~Berdibek. Reproduced with permission.}
  \label{fig:abdikamalov.kinenergy}
\end{figure}

Abdikamalov et al.~\cite{abdikamalov:16} considered incoming field of
vorticity and entropy waves. When vorticity and/or entropy waves hit a
shock, they generate a downstream field of vorticity, entropy, and
acoustic waves. Abdikamalov et al.~\cite{abdikamalov:16} showed that the
kinetic energy of fluctuations in the post-shock region is almost
entirely due to vorticity waves, with acoustic waves contributing only
$\lesssim 2\%$ of the total kinetic energy. They found that the total
kinetic energy of turbulent fluctuations is amplified by a factory of
${\sim} 2$ at shock crossing. The angular component is amplified by a
factor of ${\sim} 3$, while the radial component does not undergo
significant amplification (cf. Fig~\ref{fig:abdikamalov.kinenergy}).

\subsubsection{Impact on the Explosion Condition}
The impact of the progenitor asphericities on the explosion condition has
been estimated analytically using the concept of the critical (i.e.,
mininum) neutrino luminosity necessary to drive an explosion
\cite{mueller:15, mueller:16, abdikamalov:16}; but, see: Nagakura et
al.~\cite{nagakura:13} for a possible alternative approach. In the
absence of non-radial motion in the gain region, the critical luminosity
can be estimated as \cite{janka:12a}
\begin{equation}
  (L_\nu E_\nu^2)_\mathrm{crit} \propto (\dot M
  M)^{3/5}\ r_\mathrm{gain}^{-2/5},
\end{equation}
which is approximately equivalent to condition
$\tau_\mathrm{adv}/\tau_\mathrm{heat} \gtrsim 1$, where
$\tau_\mathrm{adv}$ and $\tau_\mathrm{heat}$ are the timescales of
advection and heating in the gain region, and $r_{\rm gain}$ is the
radius at which the net neutrino heating becomes positive
\cite{janka:12a}. The effect of post-shock turbulence can be included as
an isotropic pressure contribution (cf.\ Eq.~\ref{eq:total.pressure}),
which leads to a lower critical luminosity \cite{mueller:15}:
\begin{equation}
\label{eq:critlum}
  (L_\nu E_\nu^2)_\mathrm{crit} \propto (\dot M
  M)^{3/5}\ r_\mathrm{gain}^{-2/5}\left(1+\frac{4\langle {\cal
  M'}_2^2\rangle}{3}\right)^{-3/5},
\end{equation}
where $\langle {\cal M'}_2^2 \rangle$ is the average Mach number of
the turbulent flow in the post-shock region. Pre-shock
fluctuations cross the shock and contribute to ${\cal M'}_2$ in the
post-shock region. The reduction of the critical luminosity due to
this ``direct injection'' of kinetic energy can be estimated as
\cite{abdikamalov:16}
\begin{equation}
  \label{eq:critlum1}
  \frac{\Delta L_\mathrm{crit}}{L_\mathrm{crit}} \simeq \frac{4}{5}
  \frac{E'_\mathrm{a,2}}{E'_\mathrm{a,1}} \frac{\langle c_\mathrm{s,1}^2
  \rangle}{\langle c_\mathrm{s,2}^2 \rangle} \langle {\cal M'}_1^2
  \rangle,
\end{equation}
where $\langle c_\mathrm{s,1}^2 \rangle$ and $\langle c_\mathrm{s,2}^2
\rangle$ are the sound speeds in the pre-shock and post-shock regions. In
the absence of turbulent dissipation, the ratio
$E'_\mathrm{a,2}/E'_\mathrm{a,1}$ of the turbulent kinetic energies in
the immediate post-shock and pre-shock regions can be obtained from the
LIA formalism \cite{abdikamalov:16}. For typical parameters of the
accretion shock, and assuming anisotropic turbulence with
$R_{rr}=R_{\theta\theta}+R_{\phi\phi}$, Eq.~(\ref{eq:critlum1}) yields
\begin{equation}
  \label{eq:critlum2}
  \frac{\Delta L_\mathrm{crit}}{L_\mathrm{crit}} \simeq 0.6 \langle {\cal M'}_1^2 \rangle,
\end{equation}
The turbulent Mach number in nuclear burning shells is ${\sim} 0.1$
(e.g.,~\cite{mueller:16,collins:2017a}). The conservation of angular
momentum dictates that, in the absence of turbulent dissipation and
nuclear burning, the Mach number of non-radial fluctuations should
increase as $\propto r^{(3\gamma-7)/4}$ during accretion
\cite{kovalenko:98,lai:00}. Accordingly, if convective shells accrete
from ${\sim} 1500\,\mathrm{km}$ toward the shock at
${\sim} 200\,\mathrm{km}$, the turbulent Mach number increases by
a factor of ${\sim} 4.5$ to ${\sim} 0.45$. Compared to the case with no
pre-collapse perturbations, this leads to a reduction of the critical
luminosity by ${\sim} 12\%$.

Recent 3D neutrino-hydrodynamics simulations of the $18M_\odot$
progenitor model infer a reduction of ${\sim} 20\%$ in the critical
luminosity \cite{mueller:17}, which is significantly larger than the
above estimate. This suggests that the direct injection of kinetic
energy plays a sub-dominant role in triggering an explosion
\cite{mueller:15}. An additional contribution can come from buoyant
driving of convection in the post-shock region.

An alternative approach that approximately includes this effect was
presented by \cite{mueller:16}. Their calculation is based on the
observation that pre-collapse perturbations with Mach number ${\cal M}$
generate density perturbations with relative amplitude of ${\sim} {\cal
M}$ by the time they reach the shock. After crossing the shock, these
density perturbations become buoyant and seed turbulent convection in the
gain region. The reduction of the critical luminosity can be estimated as
\cite{mueller:16}
\begin{equation}
\frac{\Delta L_\mathrm{crit}}{L_\mathrm{crit}} \simeq 0.47 \frac{{\cal
    M}}{\ell \eta_\mathrm{acc} \eta_\mathrm{heat}},
\end{equation}
where $\ell$ is the peak angular wave number of the perturbations,
$\eta_\mathrm{heat}$ is the neutrino-heating efficiency,
$\eta_\mathrm{acc}$ is the efficiency of conversion of accretion energy
into electron-flavor neutrino emission. Recently, Huete et al.
\cite{huete:17} pointed out the importance of entropy perturbations
generated by the interaction of the supernova shock with vorticity waves
originating from nuclear burning shells. These perturbations are
associated with density fluctuations, which become buoyant and contribute
to the turbulent convection in the  gain region. Depending on problem
parameters, the resulting reduction in the critical luminosity was
estimated to be
\begin{equation}
\frac{\Delta L_\mathrm{crit}}{L_\mathrm{crit}} \simeq (0.68 - 0.96)
\times \frac{{\cal M}}{\ell \eta_\mathrm{acc} \eta_\mathrm{heat}}.
\end{equation}
For typical values of $\eta_\mathrm{heat} \simeq 0.1$, and
$\eta_\mathrm{acc} \simeq 2$, this implies a ${\sim} 17- 24\%$ reduction of the
critical luminosity for ${\cal M} \simeq 0.1$ and $\ell=2$. This is
roughly in line with the results of 3D neutrino-hydrodynamics
simulations of \cite{mueller:17}.

\subsubsection{Normal Mode Analysis.}
Takahashi et al.~\cite{takahashi:16} studied the influence of upstream
perturbations on convective and SASI modes. Their model is based on
the solution of linearized hydrodynamic equations in the post-shock
region with a boundary condition at the shock that mimics pre-shock
perturbations. They considered a stationary spherically symmetric
background model with neutrino heating. They applied perturbations of
the specific form
\begin{eqnarray}
  \frac{\delta \rho}{\rho} &=& \sin (\omega_\mathrm{up}t+\phi), \\
  \frac{\delta \upsilon}{\upsilon_{r}} &=& -0.5 \sin (\omega_\mathrm{up}t+\phi), \\
  \frac{\delta \epsilon}{\epsilon} &=& \sin (\omega_\mathrm{up}t+\phi),
\end{eqnarray}
Here, $\omega_\mathrm{up}$ is the frequency, $\phi$ is the phase of the
perturbation, and $\epsilon$ is the specific internal energy. The
linearized equations for the perturbation field are then solved using the
Laplace transform, which allowed them to calculate the frequencies and
growth rates of the SASI and convection modes under the influence of
these perturbations. Takahashi et al.~\cite{takahashi:16} systematically
studied the dependence on frequency and phase of the perturbations and
showed that resonant amplification of SASI and convection can occur when
both the frequency and the growth rate of upstream perturbations match
those of the intrinsic SASI and convective modes, respectively. Due to
this rather stringent requirement, such resonances are unlikely to occur
in CCSNe. However, they found that the amplitudes of intrinsic modes
become larger by a factor of $\lesssim 10$ when the frequency of the
upstream perturbations is close to that of the intrinsic SASI mode or to
the growth rate of convection. This suggests that accreted perturbations
might efficiently drive SASI and/or convection to nonlinear amplitudes.

\section{Protoneutron Star Convection}
\label{sec:pns}

The gain region is not the only area with convective overturn active
after core bounce. Immediately after neutrino shock breakout, the
weakening shock leaves behind a negative entropy gradient, which rapidly
develops convection \cite{epstein:79, burrows:87c, burrows:88b,
bethe:90}. However, this initial convective overturn subsides within
${\sim}10$ ms \cite{burrows:93b}, as the unstable entropy gradient is
erased by neutrino heating and convective mixing.

\begin{figure}
  \begin{center}
    \includegraphics[width=0.5\textwidth]{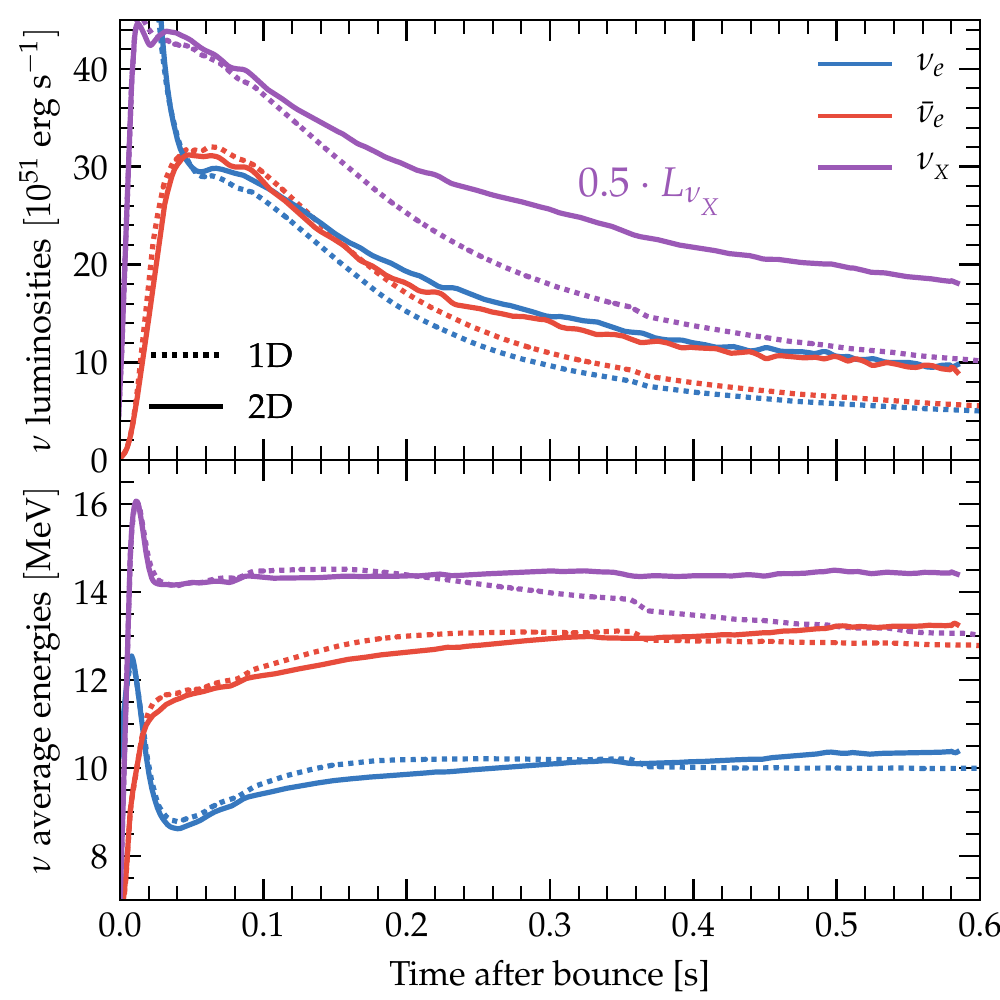}
  \end{center}
  \caption{Neutrino luminosity (top panel) and average energies (bottom
  panel) at 10,000 km as a function of time for the ONeMg
  electron-capture supernova progenitor of Nomoto \cite{nomoto:82,
  nomoto:84} computed by Radice et al.~\cite{radice:17b} with their
  ``Baseline'' setup. Here, ``$\nu_{{}_X}$'' denotes the sum of all the
  heavy-lepton neutrinos. Since this progenitor model explodes in 1D and
  2D, the neutrino luminosities in both cases are only due to the cooling
  \ac{PNS} and there is no contribution from the accretion luminosity.
  Consequently, the differences between 1D and 2D are attributable to the
  effect of PNS convection.}
  \label{fig:radice.pns}
\end{figure}

Later, neutrino losses from the outer layers of the \ac{PNS} establish
negative entropy and lepton number gradients at densities above
${\sim}10^{13}\ {\rm g}\, {\rm cm}^{-3}$. A second, much longer, period
of convective overturn develops because of the presence of these
gradients~\cite{bethe:90, burrows:93b, bhf:95, mezzacappa:98b, buras:06b,
dessart:06pns}.  The region of the \ac{PNS} that is unstable to
convection depends on the nuclear EOS, which can be seen from the
thermodynamic derivatives appearing in the Brunt-V\"ais\"al\"a frequency
(see Equation \ref{eq:omega.bv}). In general, $(\partial p / \partial
s)_{\rho,Y_e}$ is positive, so negative entropy gradients drive
instability. On the other hand, $(\partial p / \partial Y_e)_{\rho,s}$
can be negative under the conditions encountered in the outer layers of
the \ac{PNS}. Here, the pressure is mainly determined by high-density,
interacting neutron gas and the degenerate electron gas. If the neutron
gas provides the dominant contribution, reducing the electron fraction
increases the pressure. This implies the negative lepton gradients in the
outer layers of the \ac{PNS} can have a stabilizing influence for some
EOSs and the convectively unstable region will depend on the nuclear EOS.
This second phase of \ac{PNS} convection can persist for many seconds and
leave a detectable signature in the neutrino signal from the cooling
\ac{PNS} \cite{roberts:12}.

Early 2D calculations with energy-integrated neutrino transport
\cite{burrows:93b} suggested that the convective transport of energy and
lepton number from the \ac{PNS} core toward the neutrinospheres could
play an important role in the dynamics of the explosion, since it
resulted in an increase of the neutrino luminosities by up to a factor
two. However, the first 2D simulations with energy-dependent neutrino
transport found a very different picture \cite{buras:06b, dessart:06pns}.
In these models, \ac{PNS} convection was buried deep below the
neutrinospheres by large stabilizing entropy gradients interior and
exterior to it.  Accordingly, \ac{PNS} convection could only provide a
${\sim} 15\%$ to ${\sim} 30\%$ increase of the neutrino luminosities in
the first $300\ {\rm ms}$ after bounce \cite{buras:06b, dessart:06pns}.

More recently, Radice et al.~\cite{radice:17b} pointed out that, over
longer timescales, the contraction of the \ac{PNS} proceeds to the point
that its surface comes into contact with the \ac{PNS} convection region.
Afterward, they found differences in the neutrino luminosities as large
as a factor two between their 1D and 2D calculations, which they
attributed to \ac{PNS} convection. Fig.~\ref{fig:radice.pns} shows the
neutrino luminosities and average energies from their simulation of the
$8.8\ M_\odot$ (zero-age main sequence) progenitor from Nomoto
\cite{nomoto:82, nomoto:84}. Since this progenitor explodes quickly
(${\sim}50$ milliseconds after bounce) even in 1D calculations, the
differences in the neutrino luminosities are inescapably related to
\ac{PNS} convection and are not due to differences in the accretion flow.
The figure demonstrates ${\sim}2\times$ boost of the luminosities, as
well as a slight increase in the average neutrino energies. A similar
enhancement of the heavy-lepton neutrino luminosities was also found by
O'Connor \& Couch~\cite{oconnor:15b}. However, they did not find a
correspondingly large boost of the electron-type neutrino luminosity,
presumably because the smaller luminosities in their 1D calculations were
offset by the larger accretion luminosities.

Other consequences of \ac{PNS} convection are a reduced contraction
rate of the \ac{PNS} \cite{buras:06b, radice:17b} and the launching of
gravity waves in the stably stratified region above it
\cite{dessart:06pns}. The first effect is mostly negative for the
prospects of the explosion, since it results in a reduction of the
average neutrino energies by several percent. This reduction, in turn,
reduces the coupling between the neutrinos and the matter behind the
shock. The second effect is, instead, positive, since gravity waves
transport energy out of the core and release it near the surface of the
\ac{PNS}. This energy is subsequently radiated as neutrinos and, in part,
reabsorbed in the gain layer behind the shock.

The Ledoux-stable region above the inner convection layer received
significant attention before the advent of multidimensional simulations
with spectral neutrino transport. Wilson and Mayle~\cite{wilson:88,
wilson:93} suggested that a doubly diffusive instability might operate in
this region. They argued that if a parcel of neutron rich matter is
displaced toward the center of the \ac{PNS}, then it would thermally
equilibrate on a very short timescale. However, because of the net
cancellation of the interaction with electron and anti-electron
neutrinos, this material would still be more neutron rich than the
surroundings. As a result of the smaller electron pressure, the fluid
blob would keep on sinking. This process is called neutron fingering in
analogy with the salt fingering observed in Earth's oceans
\cite{schmitt:95}. Convective overturn would then rapidly develop and
transport lepton number and heat from the core of the \ac{PNS} all the
way to the neutrinospheres. This would significantly boost the neutrino
luminosities. Indeed, their 1D models with an effective prescription of
the neutron-finger instability successfully exploded \cite{wilson:93}.

However, Bruenn \& Dineva~\cite{bruenn:96} later pointed out that a
radially displaced fluid element below the neutrinospheres would reach
equilibration mostly because of the emission of low energy electron-type
neutrinos. The reason is, on the one hand, that heavy-lepton neutrinos
are not effective at transporting heat and, on the other hand, that high
energy electron-type neutrinos would be trapped due to their small mean
free path. Consequently, the displaced fluid element would achieve
lepton-number re-equilibration before cooling down to the ambient
temperature, and the neutron-finger instability should not develop. This
has since been confirmed by multidimensional simulations, which found no
evidences for the neutron finger or other doubly diffusive instabilities
\cite{bruenn:04} in the \ac{PNS} \cite{buras:06b, dessart:06pns}.

\section{Magnetic Effects}
\label{sec:mhd}

Magnetic fields can play an important role in the \ac{CCSN} explosion
mechanism. In rapidly rotating progenitors, they act as an agent powering
jet-like outflows by extracting rotational energy,
e.g.,~\cite{burrows:07b,takiwaki:11,moesta:14b}. The necessary
magnetar-strength field (${\sim}10^{15}\ {\rm G}$) can be obtained via
flux-compression and winding from strongly magnetized progenitor
cores~\cite{wheeler:02} or via
magnetoturbulence~\cite{akiyama:05,obergaulinger:09,moesta:15} in the
shear layer surrounding the \ac{PNS}. While many studies have focused on
magnetic fields in rapidly rotating progenitors, the impact of magnetic
fields in slowly or non-rotating progenitors has so far received limited
attention in the literature.

Endeve et al.~\cite{endeve:10, endeve:12} investigated magnetic field
amplification by the SASI in 2D axisymmetric and 3D simulations.  They
found that SASI-driven turbulence in the post-shock region amplifies
magnetic field exponentially exponentially (i.e., by multiple orders of
magnitude) and can account for neutron star magnetic field of ${\sim}
10^{14}\, \mathrm{G}$ or more in non-rotating progenitors. However, while
the magnetic field amplification impacted the local flow properties in
their simulations, they did not find a modification of the global
dynamics.

Obergaulinger et al.~\cite{obergaulinger:14} studied magnetic
amplification and evolution of non-rotating magnetized \acp{CCSN} via 2D
axisymmetric simulations. They considered a $15\, M_{\sun}$-progenitor
augmented with a seed magnetic fields, the strength of which they varied.
They found that convection and the SASI amplify magnetic
fields in the gain region up to a factor ${\sim} 5$, which they explained
in terms of the ratio between the advection and the eddy-turnover times.
As a consequence of this modest amplification factor their weakly
magnetized models -- with pre-collapse fields smaller than $10^{12}\ {\rm
G}$ -- evolved very similarly to purely hydrodynamical models. Only in
their $10^{12}\, \mathrm{G}$ model did the magnetic field energy reach
equipartition with the kinetic energy, thereby enabling it to trigger a
considerably earlier explosion.

Summarizing, extant studies on magnetized \acp{CCSN} are inconclusive
concerning the role that magnetic fields can play in the explosion of
slowly- or non-rotating progenitors. On the one hand, global qualitative
changes in the dynamics have been reported only for very large
pre-collapse fields. On the other hand, a quantitative impact on the flow
is present even for realistic seed fields. Moreover, the simulations that
have been carried out so far only establish a lower bound for the
amplification factor of the magnetic fields. Axisymmetric simulations
significantly restrict the possibilities for magnetic field
amplification, as illustrated by the anti-dynamo theorem
\cite{cowling:1933}, see also \cite{brandenburg:05}. In addition, 3D
studies are limited in resolution and may have not reached sufficiently
high effective Reynolds numbers to fully capture the magnetic field
amplification. Future studies that investigate magnetic field
amplification and their potential to alter the flow properties in 3D are
needed to understand their role and impact for slowly or non-rotating
\ac{CCSN} progenitors.

\section{Conclusions}
\label{sec:conclusions}

Turbulence is an important component of the \ac{CCSN} explosion
mechanism. Turbulent convection during the last stages of nuclear burning
creates perturbations that are then frozen in the collapsing gas after
the unset of core collapse. These perturbations can seed fluid
instabilities, such as the \ac{SASI} or neutrino-driven convection, as
they are accreted by the shock. The development of strong \ac{SASI}
and/or neutrino-driven convection, in turn, creates favorable conditions
for the explosion. Convection inside the \ac{PNS} can boost the neutrino
luminosity and might in some cases contribute to shock revival. Finally,
magnetic fields are expected to be amplified by convection and/or by
other MHD instabilities, such as the \ac{MRI}, and might play an
important role in the explosion.

Turbulence may also be important in understanding and interpreting observations of CCSNe.
This is particularly true for CCSN remnants such as Cassiopeia A \cite{lopez:18}.
In Cas A, recent observations suggest the presence of radioactive nickel-heated bubbles expanding in the remnant, the structure of which may be reflective of the turbulence in the explosion mechanism itself \cite{milisavljevic:15}.
Additionally, the presence of high-velocity knots at large radius in the remnant could also be the result of turbulent structures in the mechanism that have become ballistic within the expanding remnant \cite{fesen:16}.
Observational evidence for turbulence in the CCSN mechanism also comes from complex spectral features in several objects.
In the case of SN 1987A, detailed analysis of He I lines indicates the turbulent mixing of radioactive nickel to large velocities \cite{fassia:99}.
The so-called ``Bochum'' event in SN 1987A \cite{hanuschik:88} may also be evidence for turbulent mixing in the explosion \cite{wang:02}.
Multiply peaked O lines in various stripped envelope CCSNe entering the nebular phase may also imply the action of turbulence in the inner workings of the CCSN mechanism \cite{modjaz:08a}.

Many of these important parts of the \ac{CCSN} explosion mechanism puzzle
have been pieced together in the past few years. However, this progress
has been driven either by specialized/idealized simulations that
addressed only one of these aspects in isolation, or by means of
simulations that ostensibly included all of the relevant physics, but
lacked the numerical resolution necessary to capture all scales of the
problem. Solving the \ac{CCSN} explosion mechanism problem will
necessarily require a way to bridge the gap between these two approaches.
A possible way forward might be the development of subgrid-scale models
for the effective treatment of turbulence in global simulations. This
will be an object of our future work.

\ack
The authors acknowledge Adam Burrows, J\'er\^ome Guilet, Roland Haas,
Thierry Foglizzo, Bernhard M\"uller, Jeremiah W.~Murphy, and Evan
O'Connor for insightful discussion on the explosion mechanism of
core-collapse supernovae, and the anonymous referee for valuable
suggestions that improved the quality of this manuscript.  DR gratefully
acknowledges support from the Schmidt Fellowship, the Sherman Fairchild
Foundation and the Max-Planck/Princeton Center (MPPC) for Plasma Physics
\mbox{(NSF PHY-1523261)}. CDO is partially supported by NSF grant CAREER
PHY-1151197.
SMC is supported by the U.S. Department of Energy, Office of Science, Office of Nuclear Physics, under Award Numbers DE-SC0015904 and DE-SC0017955 and the Chandra X-ray Observatory under grant TM7-18005X.
Work presented in this review benefitted from computer time
allocations at NSF/NCSA Blue Waters (PRAC ACI-1440083), at the National
Energy Research Scientific Computing Center (project m152), a DOE Office
of Science User Facility supported by the Office of Science of the U.S.
Department of Energy under Contract No.\ DE-AC02-05CH11231, and on the
Texas Advanced Computing Center Stampede cluster under NSF XSEDE
allocation TG-PHY100033.
Simulations described herein where completed with computer time provided by the Innovative and Novel Computational Impact on Theory and Experiment (INCITE) program. This research used resources of the Argonne Leadership Computing Facility, which is a DOE Office of Science User Facility supported under Contract DE-AC02-06CH11357.


\bibliography{%
bibliography/bh_formation_references,%
bibliography/gw_references,%
bibliography/sn_theory_references,%
bibliography/grb_references,%
bibliography/nu_obs_references,%
bibliography/methods_references,%
bibliography/eos_references,%
bibliography/NSNS_NSBH_references,%
bibliography/stellarevolution_references,%
bibliography/nucleosynthesis_references,%
bibliography/gr_references,%
bibliography/nu_interactions_references,%
bibliography/sn_observation_references,%
bibliography/populations_references,%
bibliography/pns_cooling_references,%
bibliography/spectral_photometric_modeling_references,%
bibliography/cs_hpc_references,%
bibliography/numrel_references,%
bibliography/mhd_references,%
bibliography/gw_data_analysis_references,%
bibliography/gw_detector_references,%
bibliography/radiation_transport_references,%
bibliography/fluid_dynamics_references,%
bibliography/cosmology_references,%
bibliography/stellar_oscillations_references.bib,%
bibliography/privatebibs/radice}

\acrodef{AMR}{adaptive mesh-refinement}
\acrodef{BH}{black hole}
\acrodef{BHNS}{black-hole neutron-stars}
\acrodef{BNS}{binary neutron stars}
\acrodef{CCSN}{core-collapse supernova}
\acrodefplural{CCSN}[CCSNe]{core-collapse supernovae}
\acrodef{CMA}{consistent multi-fluid advection}
\acrodef{HMNS}{hypermassive neutron star}
\acrodef{EM}{electromagnetic}
\acrodef{EOB}{effective-one-body}
\acrodef{EOS}{equation of state}
\acrodefplural{EOS}[EOS]{equations of state}
\acrodef{GR}{general relativistic}
\acrodef{GRHD}{general-relativistic hydrodynamics}
\acrodef{GW}{gravitational wave}
\acrodef{MHD}{magnetohydrodynamic}
\acrodef{MRI}{magnetorotational instability}
\acrodef{NR}{numerical relativity}
\acrodef{NS}{neutron star}
\acrodef{PNS}{protoneutron star}
\acrodef{SASI}{standing accretion shock instability}
\acrodef{SGRB}{short gamma-ray burst}
\acrodef{SN}{supernova}
\acrodefplural{SN}[SNe]{supernovae}
\acrodef{SNR}{signal-to-noise ratio}

\end{document}